\documentclass[aps,superscriptaddress,twocolumn,pre,longbibliography,floatfix]{revtex4-1}

\usepackage{amsmath,amsthm,amsfonts,amssymb,bm,graphicx,color,mathpazo,times, braket}
\usepackage[colorlinks={true}, citecolor={blue}, filecolor={blue}, linkcolor={blue}, urlcolor={blue}]{hyperref}
\usepackage[caption=false]{subfig}

\usepackage{soul}

\allowdisplaybreaks

\begin{document}
\preprint{APS/123-QED}
\title{Enantiomer detection via Quantum Otto cycle}
\author{Mohsen Izadyari}
    \email[ ]{mizadyari18@ku.edu.tr}
    \affiliation{Department of Physics, Ko\c{c} University, 34450 Sar\i yer, Istanbul, T\"urkiye}
\author{M. Tahir Naseem}
    \email[ ]{mnaseem16@ku.edu.tr}
    \affiliation{Department of Physics, Ko\c{c} University, 34450 Sar\i yer, Istanbul, T\"urkiye}
\author{\"Ozg\"ur E. M\"ustecapl\i o\u glu}
    \email[ ]{omustecap@ku.edu.tr}
    \affiliation{Department of Physics, Ko\c{c} University, 34450 Sar\i yer, Istanbul, T\"urkiye} 
    \affiliation{T\"UB˙ITAK Research Institute for Fundamental Sciences, 41470 Gebze, T\"urkiye}
\date{\today}
\begin{abstract}
 Enantiomers are chiral molecules that exist in right-handed and left-handed conformations. Optical techniques of enantiomers detection are widely employed to discriminate between left- and right-handed molecules. However, identical spectra of enantiomers make enantiomer detection a very challenging task. Here, we investigate the possibility of exploiting thermodynamic processes for enantiomer detection. In particular, we employ a quantum Otto cycle, in which a chiral molecule described by a three-level system with cyclic optical transitions is considered a working medium. Each energy transition of the three-level system is coupled with an external laser drive. We find that the left-handed molecule works as a heat engine, while the right-handed molecule works as a thermal accelerator where the overall phase of the drives is considered as the cycle's control parameter. In addition,  both left- and right-handed molecules work as heat engines by considering laser drives' detuning as the control parameter. However, the molecules can still be distinguished because both cases' extracted work and efficiency are quantitatively very different. Accordingly, left and right-handed molecules can be distinguished by evaluating the work distribution in the Otto cycle.
\end{abstract}

\maketitle

\section{\label{sec:Intro}Introduction }
\par Quantum thermodynamics investigates the applicability of laws of thermodynamics and their possible generalizations in the realm of quantum mechanics~\cite{QTBook01,e15062100, 00107514, Goold_2016, QTBook02,AsliReview,5.0083192}. Quantum heat engines (QHEs) serve as test beds for fundamental discussions and practical applications of quantum thermodynamics. In a broad context, QHE is a thermal machine whose operation cycle requires a quantum mechanical description. A typical example is the quantum Otto cycle, where the fast isentropic compression and expansion strokes of its classical counterpart are replaced by the slow quantum adiabatic parametric processes~\cite{Intro-QHE01,Intro-QOE01,Intro-QOE02,Intro-QOE03,Intro-QOE04}. Different quantum working substances have been proposed to implement quantum Otto cycle~\cite{Intro-QHE02,Intro-QHE03,Intro-QHE04,Intro-QHE05,Intro-QHE06,Intro-QHE07}, and experimental demonstrations have been reported for spin systems~\cite{Intro-QHE08, Bouton2021,Zhang2022}. These studies revealed that a quantum Otto cycle's work output and efficiency are determined by the energy spectrum of the quantum working system. Accordingly, we ask if a quantum Otto cycle can be used as a probe to discriminate systems with identical energy spectrums but with distinct spectral changes under the same parametric processes. A specific system with such a spectral character and for which this question is particularly significant is a chiral molecule.

	Chiral molecules can have either left- or right-handed geometries that are not superimposable on their mirror images~\cite{ChiralBook01,chiral0011}. They play significant role in fundamentals and applications in biology~\cite{Intro-bio02,Intro-bio03,Intro-bio04}, physics~\cite{Intro-phys03}, and pharmacy~\cite{Intro-farm01}. The two mirror-image molecules are called enantiomers. Enantiomers exhibit identical physical and chemical properties in an achiral environment, while they can act remarkably differently when placed in a chiral environment~\cite{Intro-bio04}. Chiral molecules may exist as a mixture of enantiomers in varying proportions. Therefore, enantioseparation is a critical and challenging task in chemistry and medicine~\cite{Intro-phys04,Intro-select05,Intro-select01,Intro-select02,Intro-select03,Intro-select06,Intro-select07}.
	
	Optical enantio-separation techniques are one of the most common methods for enantiodetection based on the interference between the electric and the magnetic dipole transitions~\cite{Intro-select07,Intro-select01}.
Enantiomer-specific microwave spectroscopic methods, based on cyclic three-level systems, which can exist in chiral molecules~\cite{Intro-threelevel03,ham02} and other artificial symmetry-broken systems~\cite{Intro-cycle06}, have been examined for the enantiodetection of chiral 
molecules~\cite{Intro-threelevel01,Intro-threelevel02,Intro-threelevel03,Intro-threelevel04,Intro-threelevel05,Intro-threelevel06,Intro-threelevel07}. Three-level quantum systems with broken symmetry allows the coexistence of one- and two-photon transitions such 
that a cyclic population transfer can occur between different energy levels~\cite{Intro-cycle01,Intro-cycle02,Intro-cycle03,Intro-threelevel03,ham02}. In this paper, we consider a chiral molecule, described by a three-level system with cyclic optical transitions, as our working substance subject to a quantum Otto cycle. By calculating the work and efficiency of the engine, we investigate if left- and right-handed enantiomers can have different energetics that can be probed by the engine behavior and the performance. 

	The rest of the paper is organized as follows. In Sec.~\ref{sec:system}, we describe and shortly review the Hamiltonian model to describe a cyclic three-level chiral molecule interacting with three linearly polarized optical fields, which was
originally proposed in Ref.~\cite{Intro-threelevel03}. In Sec.~\ref{Sec:OttoCycle}, we introduce the quantum Otto cycle operation where the detuning and the phases of the optical fields are used as the control parameters in 
the quantum adiabatic strokes. We calculate the 
quantum thermodynamic work and efficiency for different enantiomers and discuss how they can be distinguished in Sec.~\ref{sec: results}. We conclude in Sec.~\ref{sec:conclusion}.
\begin{figure}
    \centering
    \includegraphics[width=\linewidth]{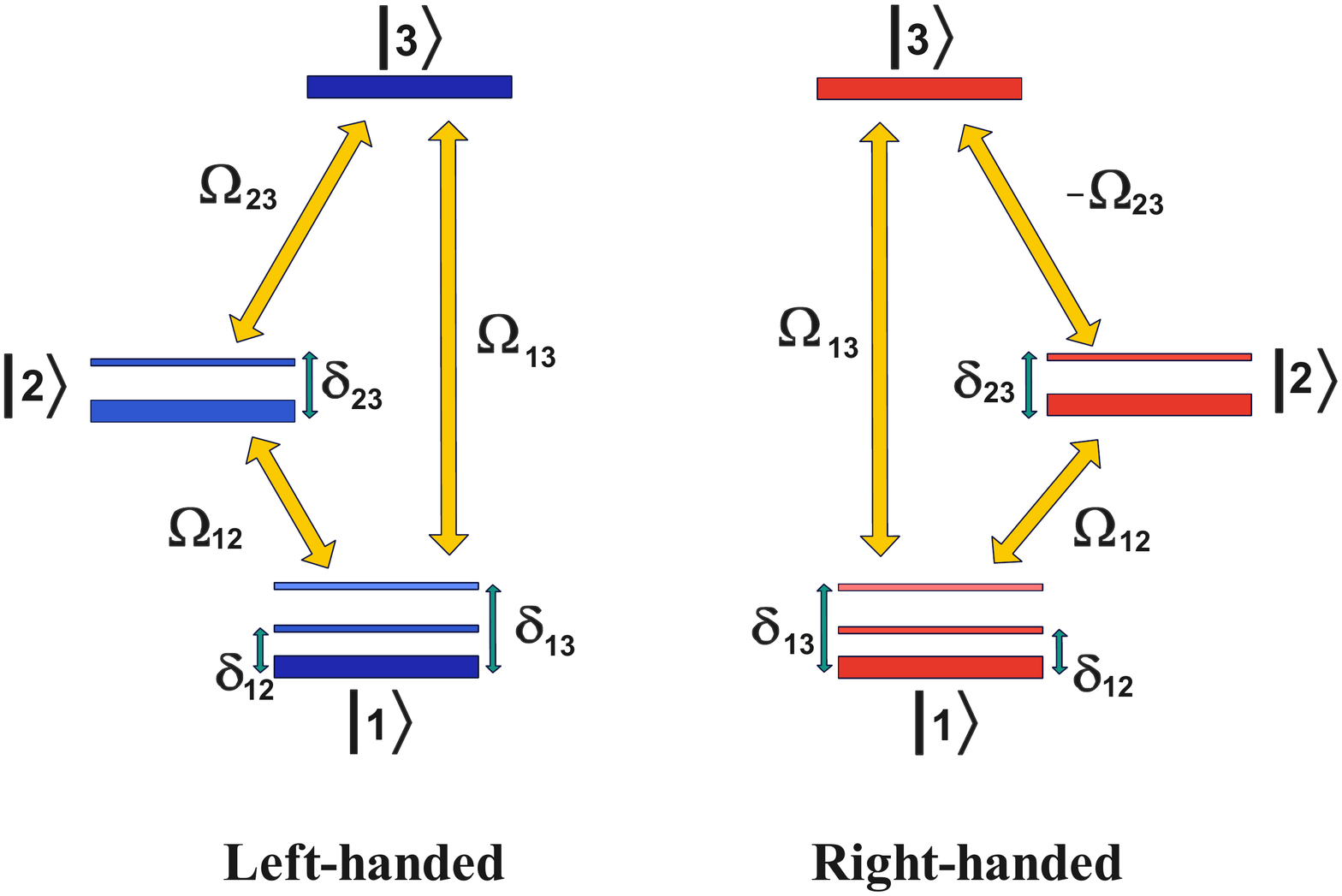}
    \caption{(Color online) A schematic illustration of a left-handed (a) and right-handed (b) chiral molecule which is described by a three-level system with cyclic optical transitions. The three-level system is coupled with three classical optical drives of Rabi frequencies $\Omega_{12}$, $\Omega_{13}$, and $\pm\Omega_{23}$, where $+\Omega_{23}$ ($-\Omega_{23}$) describes the left-handed (right-handed) molecule. The detunings of the external optical drives from energy transitions are indicated by $\delta_{ij}$.} 
    \label{fig:fig1}
\end{figure}
\begin{figure*}[ht]
    \centering
    \subfloat[]{{\includegraphics[width=0.24\textwidth]{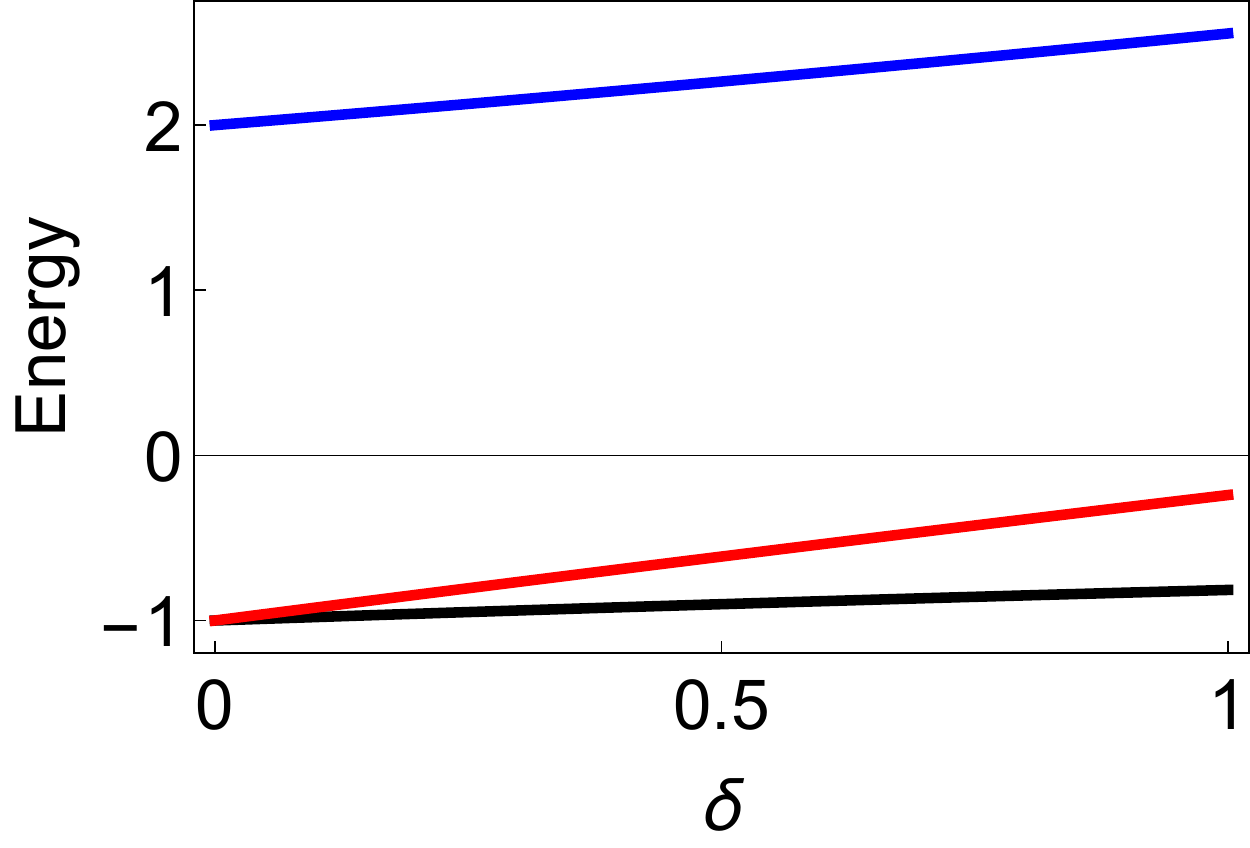} }}
    \hfil
    \subfloat[]{{\includegraphics[width=0.24\textwidth]{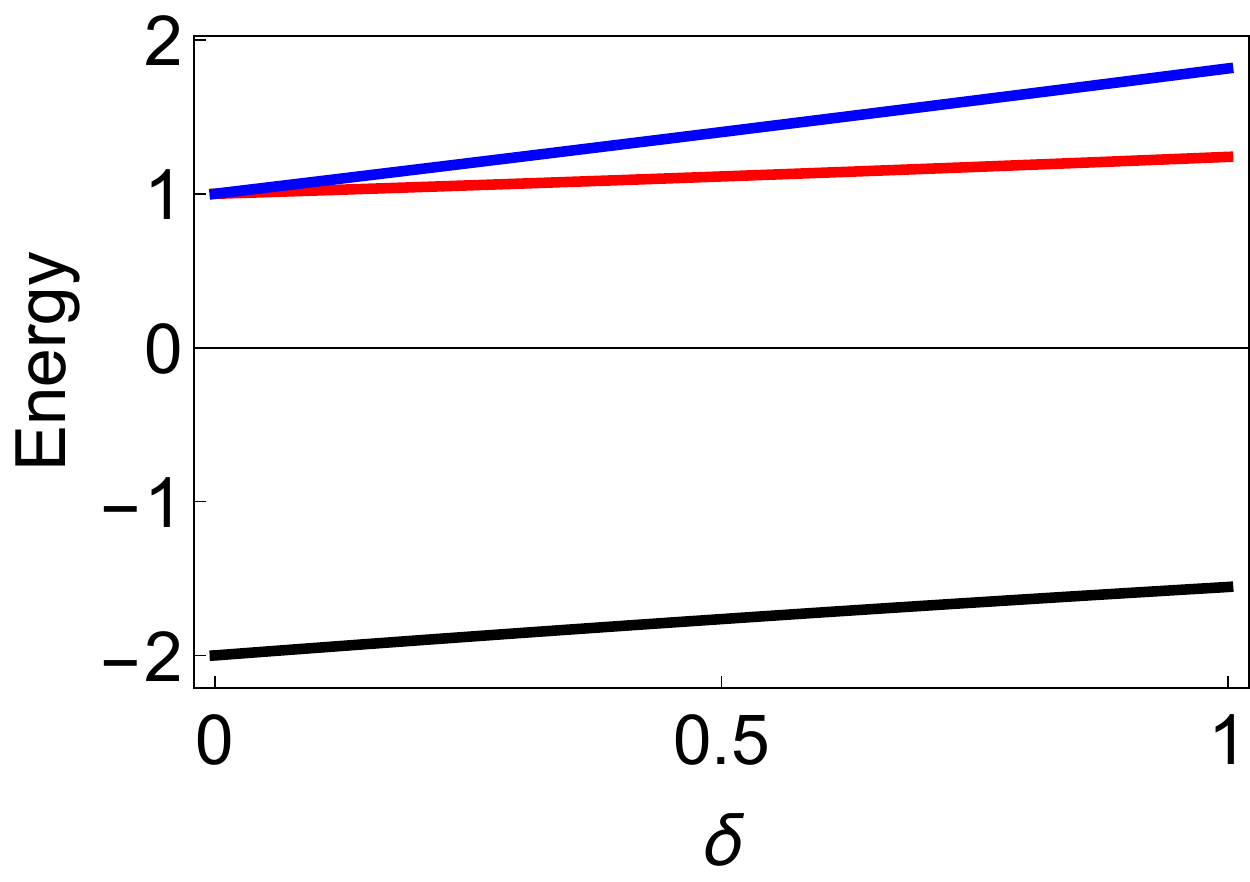} }}
    \hfil
    \subfloat[]{{\includegraphics[width=0.24\textwidth]{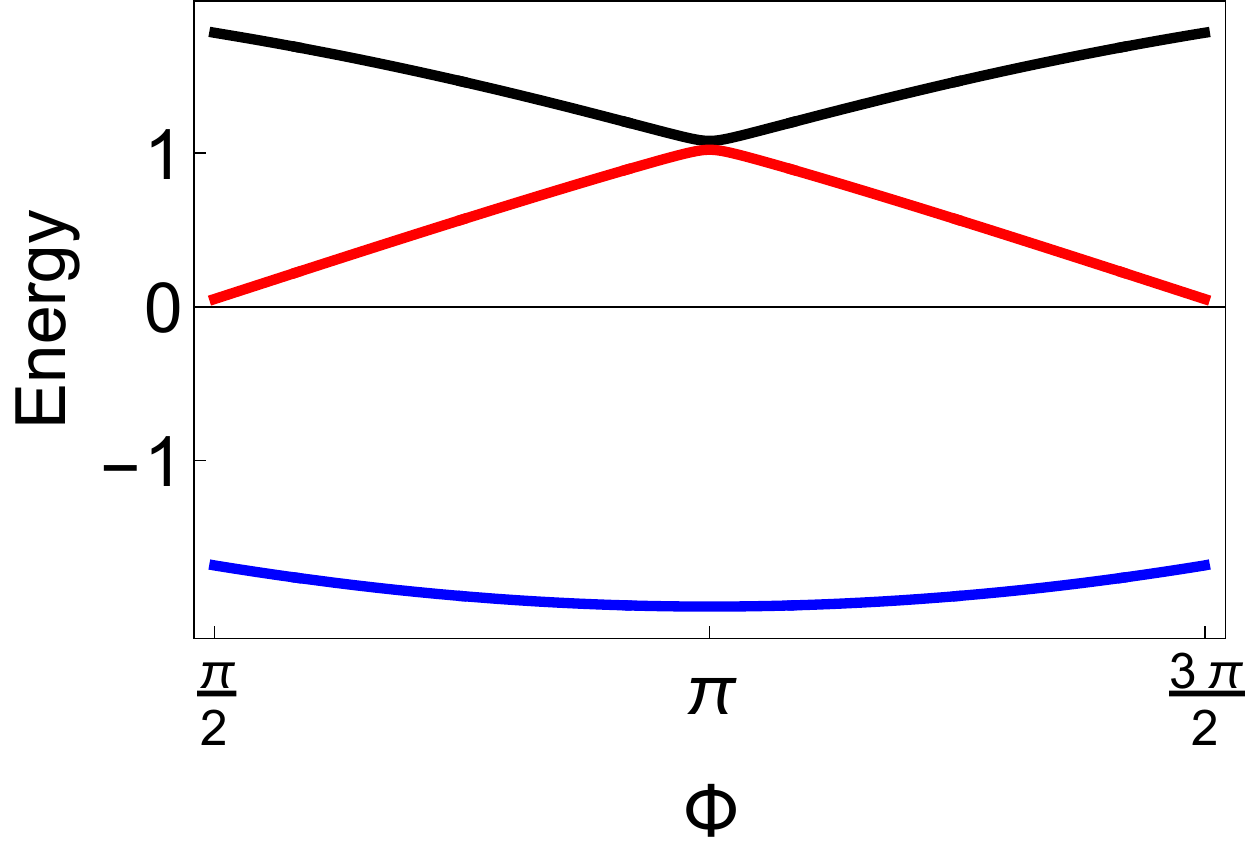} }}
    \hfil
    \subfloat[]{{\includegraphics[width=0.24\textwidth]{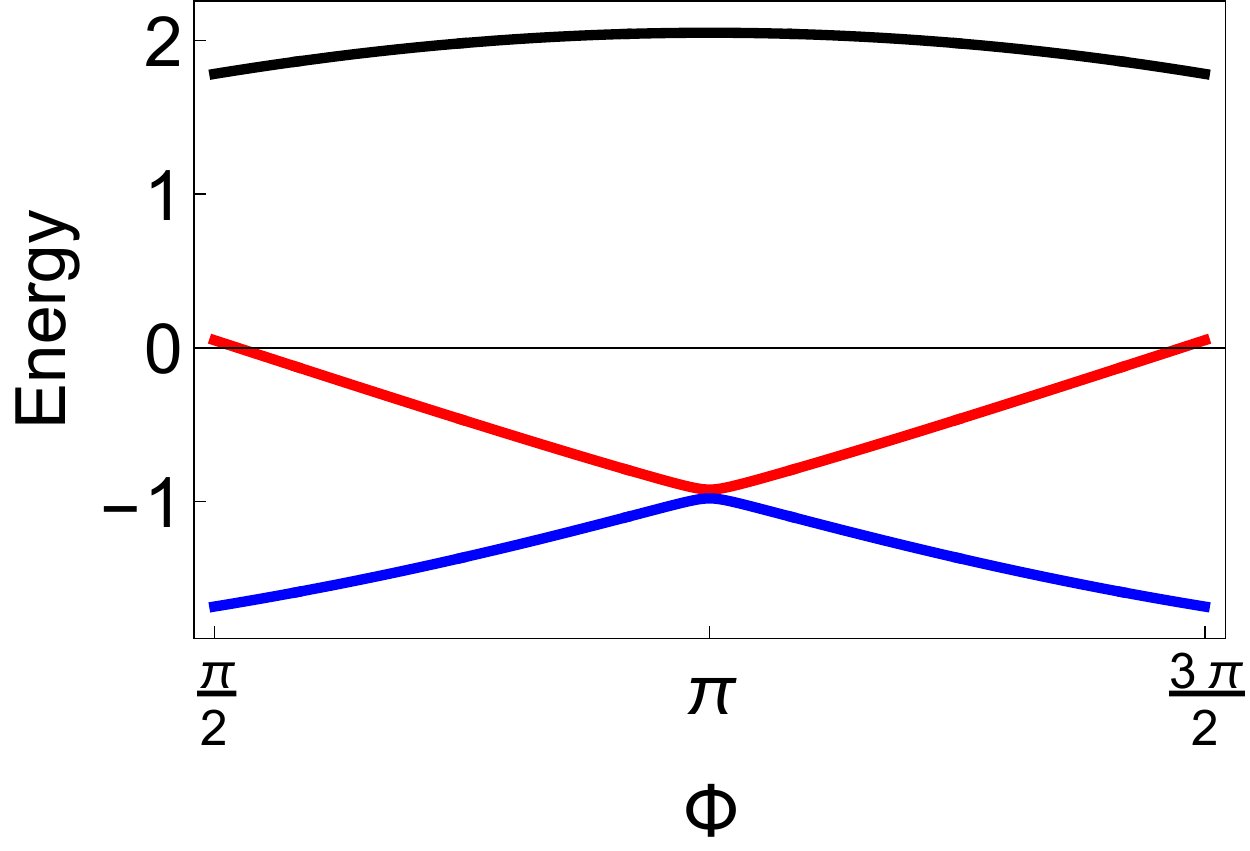} }}
    \caption{(Color online) Energy eigenvalues of the Hamiltonian $H^{\pm}_\text{int}$ for constant overall phase $\Phi = 0$ (a, b), and constant detuning $\delta = 0.1/\tau_0$ (c, d). The panels (a, c) and (b, d) represent left- and right-handed enantiomers, respectively. The eigenenergies are scaled with the rotational energy $E_0 = \hbar^2 B$ with the rotational constant $B$, that is why eigenenergies are $\pm1 E_{0}$ and $\pm2 E_{0}$ for $\Phi=\delta=0$. The eigenvalues are obtained for time-independent Hamiltonian ($\Delta=0$) and considering the unit magnitude of the Rabi frequencies $\Omega_{ij}$.}
    \label{fig:fig2}
\end{figure*}
\section{\label{sec:system}Model System}
We consider a chiral molecule that is described by a three-level system with cyclic optical transitions driven by three classical optical fields \cite{Intro-threelevel03,ham02,chiral001}, and it is shown in Fig.~\ref{fig:fig1}. The Hamiltonian of this system in the rotating wave approximation is given by~\cite{Intro-threelevel03,ham02,chiral001}
\begin{equation}
H = \sum^3_{j=1} \omega_{j}\ket{j}+\sum^3_{i>j=1}(\Omega_{ij}(t)e^{-i\omega_{ij}t}\ket{i}\bra{j}+\text{H.c.}).
\end{equation}
Here, the first term describes the energy of a three-level system with $\omega_{j}$ being the energy of the state $\ket{j}$. The second term describes the coupling of classical optical fields with the three-level atom; the frequency of the optical field is given by $\omega_{ij}$ which describes the coupling between $\ket{i}\leftrightarrow\ket{j}$ energy levels. $\Omega_{ij}(t)=\vec{\mu}_{ij}.\vec{E}_{ij}$ represents the Rabi frequency which depends on the field strength $\vec{E}_{ij}$ and associated transition dipole matrix element $\vec{\mu}_{ij}$ between the states $\ket{i}$ and $\ket{j}$. In the interaction picture, the Hamiltonian of the chiral molecule is given by
\begin{align} \label{eq:01Ham}
    H^{\pm}_\text{int} = \begin{pmatrix}
\delta_{13} & \Omega_{12} e^{i(\Delta t + \Phi)} & \Omega_{13} & \\
\Omega_{12} e^{-i(\Delta t + \Phi)} & \delta_{23} & \pm \Omega_{23}\\
\Omega_{13} & \pm \Omega_{23} & 0
\end{pmatrix}
\end{align}
Here, $\delta_{ij}$ is the detuning of the optical fields with energy levels $\ket{i}\leftrightarrow\ket{j}$, and $\Delta := \delta_{12}-\delta_{13}+\delta_{23}$. In addition, we define
overall phase $\Phi := \Phi_{12}-\Phi_{13}+\Phi_{23}$, where $\Phi_{ij}=\phi_{ij}+\theta_{ij}$ contains the material phase $\theta_{ij}$ and the phases of the electric fields $\phi_{ij}$~\cite{chiral001}.
We stress that the difference between the left- and right-handed molecules are expressed by the $\pm\Omega_{23}$, where the left-handed and right-handed molecules are given by positive and negative Rabi frequency $\Omega_{23}$, respectively (see Fig.~\ref{fig:fig1}).

	According to Eq.~(\ref{eq:01Ham}), the interaction Hamiltonian becomes time-independent for $\Delta = 0$~\cite{chiral001}. In the rest of the paper, we assume $\delta_{12}=\delta_{23}=\delta_{13}/2 \equiv \delta$ for which $\Delta$ becomes zero and the Hamiltonian becomes time-independent. This can be achieved by making a judicious choice on the selection of frequencies $\omega_{ij}$. All system parameters can be scaled for convenience with arbitrary energy $E_{0}$. Henceforth, we assume that the system parameters are scaled with rotational energy $E_0 = \hbar^2 B$, defined by the angular momentum operators concerning the principle molecular axes \cite{chiral001}, with the rotational constant $B$. Accordingly, time and frequencies are given in the units of $\tau_0 = \hbar/E_0$ and $1/\tau_0$.

	Fig.~\ref{fig:fig2} displays the eigenvalues of the Hamiltonian $H^{\pm}$ given in Eq.~(\ref{eq:01Ham}) as a function of the overall phase $\Phi$ and detuning $\delta$. We implement the adiabatic strokes in the quantum Otto cycle by either varying the overall phase $\Phi$ or detuning $\delta$; the adiabaticity condition requires no level crossing of the energy levels of the working medium. This is why we consider the system parameters in Fig.~\ref{fig:fig2} and the rest of the paper such that there is no level crossing in the field-dressed eigenstates, which can occur for $\Phi= 0, \pi, 2\pi$, and $\delta=0$~\cite{chiral001}. According to Fig.~\ref{fig:fig2}(c-d), the left- and right-handed enantiomer have identical energy spectrum for phases $\Phi=\pi/2$ and $\Phi=3\pi/2$. We consider these two phase values as the starting and ending points of the adiabatic strokes in the cycle in Sec.~\ref{sec: ConstantDetuning}. Although the energy spectrum is identical at the start and end of the adiabatic strokes, we will show that the Otto cycle discriminates the left- and right-handed enantiomer in the output work. Furthermore, we investigate the Otto cycle by keeping the overall phase constant and adjusting the detuning in adiabatic processes as a more practical assumption in Sec.~\ref{sec:ConstantPhase}. The next section illustrates that the different energy characteristics of enantiomers lead to different thermodynamic properties when subjected to an Otto cycle.

\section{Chiral Molecule Otto cycle}\label{Sec:OttoCycle}
\begin{figure}
    \centering
     \includegraphics[width = 0.9 \linewidth]{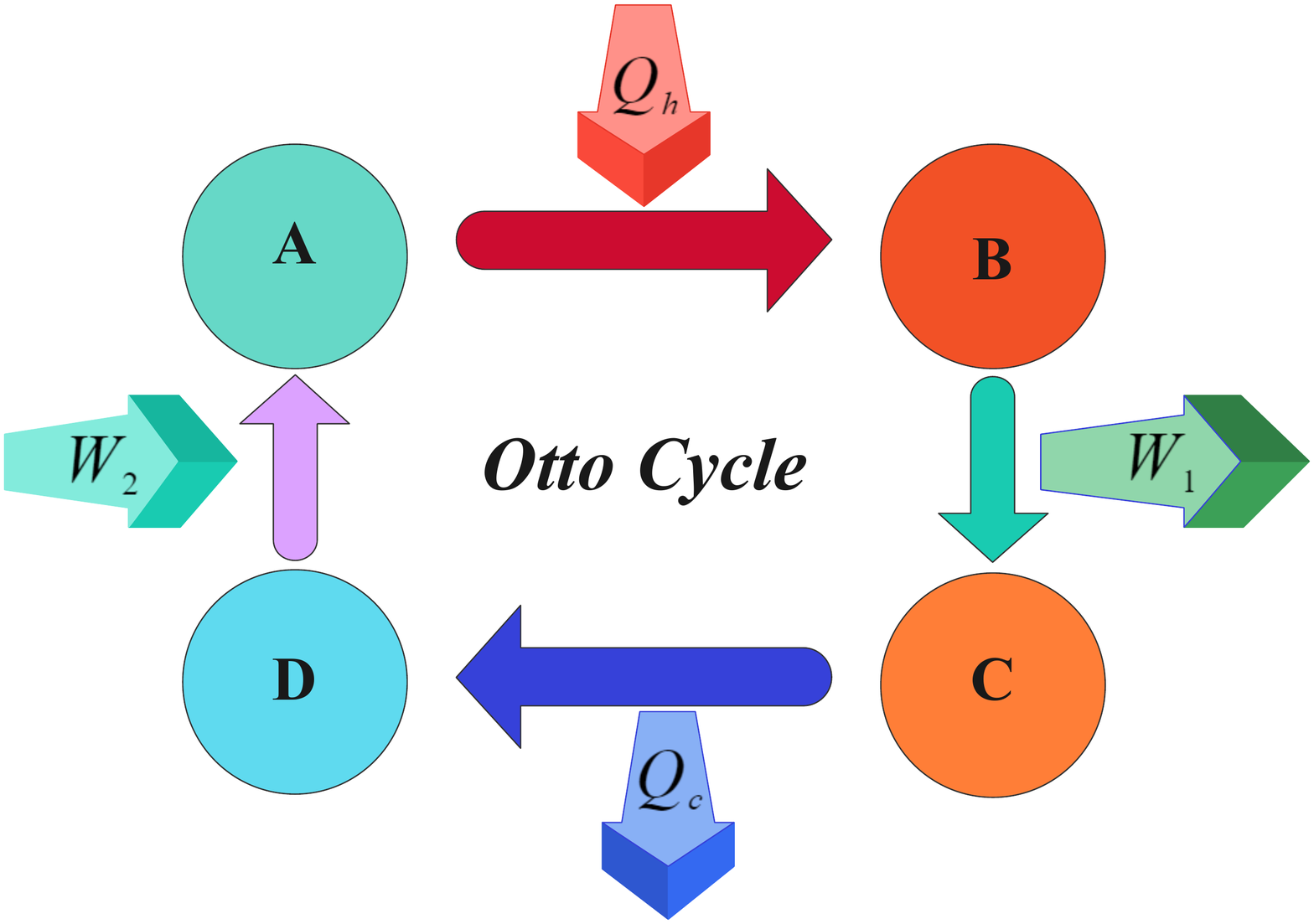}
    \caption{(Color online) A schematic description of a quantum Otto cycle consisting of two isochore strokes ($A\to B$ and $C \to D$), and two adiabatic strokes ($B\to C$ and $D \to A$). During the hot isochore $A\to B$, ${Q}_{h}$ heat is injected into the working medium from the hot bath of temperature $T_{h}$, and in the cold isochore $C\to D$, the working medium rejects ${Q}_{c}$ into the cold bath of temperature $T_{c}$. During the adiabatic expansion $B\to C$ and compression $D\to A$, work is extracted and invested on the working medium, respectively.
}
    \label{fig:fig3}
\end{figure}

A quantum Otto cycle (QOC) consists of two quantum isochoric and two adiabatic processes~\cite{QHE001}. We consider a chiral three-level system with cyclic optical transitions as a working medium of our quantum Otto cycle. A schematic illustration of the QOC is shown in Fig.~\ref{fig:fig3} where the system is in contact with a hot and cold bath in two isochoric processes indicated by $A\rightarrow B$ and $C\rightarrow D$, respectively. During the adiabatic strokes of the cycle, the  Hamiltonian changes slowly, and the system is not allowed to exchange heat with the environment. At the start of the cycle, we assume that the working medium is in thermal equilibrium with the cold thermal bath of inverse temperature $\beta_c$. The four strokes of the quantum Otto cycle are described as follows:
\par \textbf{Hot Isochore ($A \rightarrow B$)}: The first stroke is the quantum isochoric heating process in which the working medium is coupled to a hot bath of inverse temperature $\beta_h$. During this stroke, ${Q}_{h}$ heat is injected into the system, but no work is done. At the end of this stroke, we assume that the working medium comes to equilibrium with the hot thermal bath, and each energy level has an occupation probability of $P_{j}(\beta_{h})$. The exchanged heat $Q_h$ with the hot thermal bath is calculated by \cite{QHE002}
\begin{align} \label{Eq: Hot isochore}
    Q_h =  \sum_{n=1}^{3} E_n(A) (P_n (\beta_h) - P_n (\beta_c)) 
\end{align}
Here, $E_n(A)$ is $n$-th eigenenergy of system Hamiltonian $H_\text{int}$ for overall phase $\Phi_{1}$, and the population of the associated eigenstate is changed from $P_n (\beta_c)$ to $P_n (\beta_h)$. During this stage, the Hamiltonian $H^{\pm}_\text{int}$ is kept constant; accordingly, no work is done.

In general, the system may not reach a thermal state in the presence of external laser drives. Here, we assume the cyclic three-level system approximately thermalizes with the bath at the end of the isochoric stages. This approximation is reasonable for weak envelopes of the three electromagnetic drives and high temperatures of the thermal baths. Thermalization of a two-level atom in the presence of a laser drive has been discussed previously, e.g., see Eqs. (10) and (11) of Ref.~\cite{thermalization001}. 

To check the validity of the thermalization assumption, we numerically simulate the system's dynamics for the isochoric stages of the cycle. In particular, the dynamics of the reduced state $\rho$ of the three-level system during the isochoric stages is given by the master equation~\cite{QopenSystemBook,QOscully,localMasterEq}
\begin{align} \label{eq:master eq 01}
\frac{d\rho}{dt} & = -i[\hat H, \rho] + \sum_{i \neq j}\kappa_{\alpha} (\Bar{n}_{\alpha} + 1) D[\hat \sigma_{ij}] + \kappa_{\alpha} \Bar{n}_{\alpha} D[\hat \sigma_{ij}^{\dagger}].
\end{align} 
Here, $D[\hat u]=(1/2)(2\hat u \rho \hat u ^{\dagger} - \hat u^{\dagger} \rho \hat u - \rho \hat u^{\dagger} \hat u)$ refers to the Lindblad dissipator superoperators with $\hat u = \hat \sigma_{ij}$ ($i=1,2,3$). $\alpha=h,c$ denotes the hot or cold bath, respectively. In addition, $\kappa_{\alpha}$ is the system-bath coupling strength and $\bar{n}_{\alpha}$ is the average excitation of the bath.  We obtained the reduced state of the three-level system $\rho$ by
numerically solving  the master equation using Python programming language and an open source quantum optics package QuTiP~\cite{Qutip}. 

\begin{figure}
    \centering
     \includegraphics[width = 1.1\linewidth]{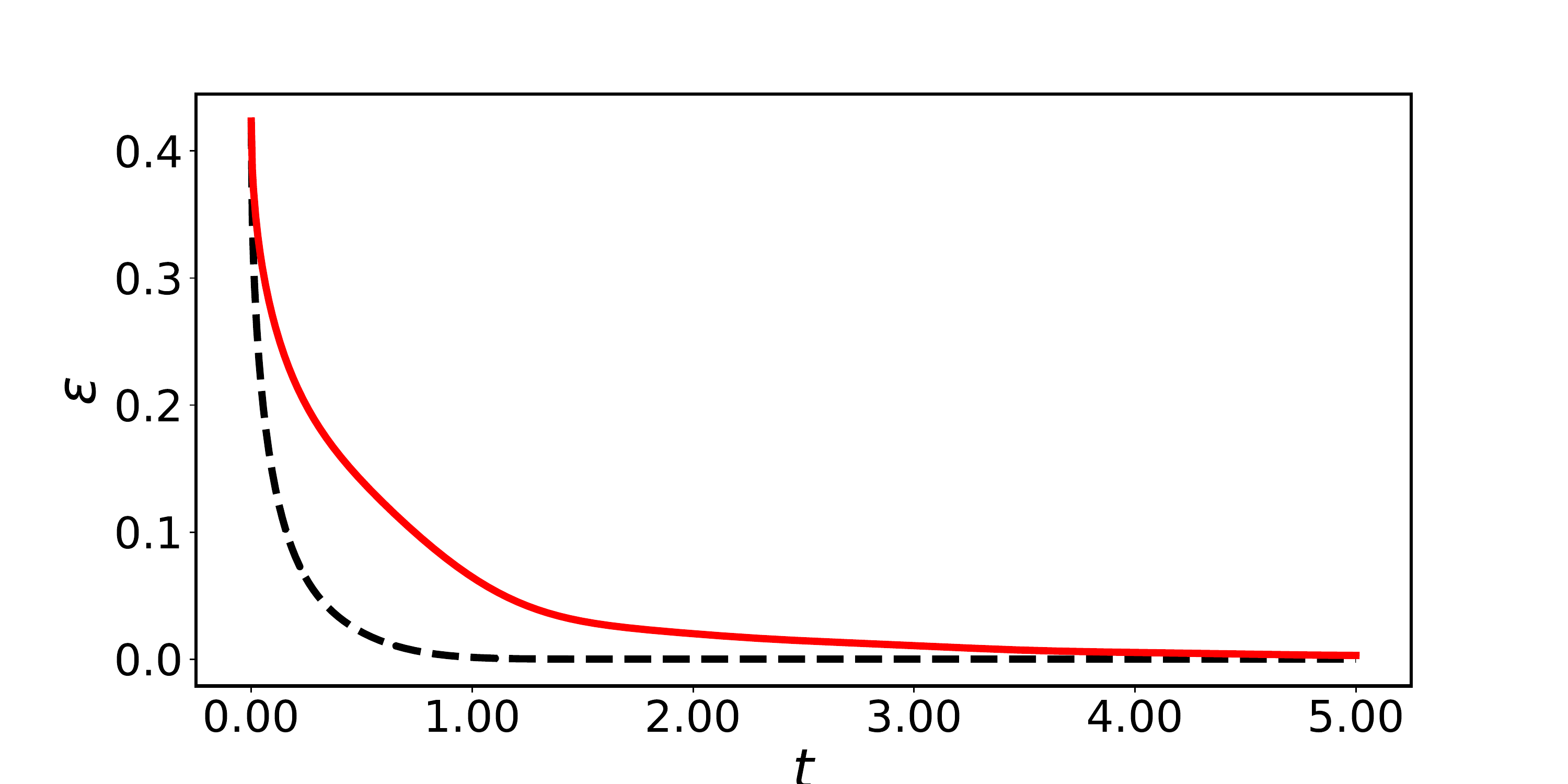}
    \caption{(Color online) The error $\varepsilon$ (Eq.~(\ref{eq:error})) in the assumption of thermalization calculated during the isochoric stages of the cycle as a function of the time (scaled by $\tau_0$). Solid and dashed lines represent the thermalization of the system during the cold and hot isochore stages of the cycle. In the long time limit, the state of the three-level system $\rho(t)$ reaches the target thermal state $\rho_\text{th}$ during the isochoric stages.}
    \label{fig:fig4}
\end{figure}

We note that a system coupled to a thermal bath at inverse temperature $\beta_h$ takes $t\to\infty$ time to reach the corresponding exact Gibbs (thermal) state~\cite{QopenSystemBook} with density matrix
	\begin{equation}
	{\rho}_\text{th} = \frac{e^{-\beta_{j}{H}_\text{s}}}{\text{Tr}[e^{-\beta_{j}{H}_
	\text{s}}]}
	\end{equation}
In order to realize a practical heat engine, we assume the system reaches the target thermal state as long as its state is at a small distance from the target state. We quantify this error by
	\begin{equation}\label{eq:error}
	\varepsilon = 1-\mathcal{F}({\rho}(t), {\rho}_\text{th}),
	\end{equation} 
here $\mathcal{F}=\text{Tr}[\sqrt{\sqrt{{\rho}}{\rho}_\text{th}\sqrt{{\rho}}}]$ is the fidelity between the state of the three-level system and its corresponding thermal state at $\beta_{j}$ during the thermalization strokes. In our numerical results, the calculated tolerance is obtained as $\varepsilon<10^{-4}$. Fig.~\ref{fig:fig4} shows the tolerance $\varepsilon$ as a function of scaled time; it is immediately evident that our assumption of thermalization is reasonable in the considered system parameters regime. It indicates that the system in each thermalization process reaches its thermal state with high fidelity ($\mathcal{F} \approx 1$).
\par \textbf{Adiabatic Expansion ($B\rightarrow C$)}: During this stage, the system is decoupled from the hot bath, and the system Hamiltonian is changed from $H(A)$ to $H(B)$. We discuss two strategies to implement the adiabatic strokes: First, varying $\Phi$ from $\Phi_1$ to $\Phi_2$ by keeping detuning constant such that $\delta > 0$. In second case, changing the detuning $\delta$ from $\delta_1$ to $\delta_2$ for constant phase $\Phi$. Work is extracted from the system during this stage of the cycle. We assume this process is slow enough that the occupation probabilities of the energy levels are unchanged to satisfy the adiabaticity condition. The adiabaticity can be ensured by modifying the system's energy over a time interval much shorter than the one needed to interact with a thermal bath \cite{adiabatic01}.

\par \textbf{Cold Isochore ($C \rightarrow D$)}: During this stage, the system is subject to either constant phase $\Phi$ and it is put into contact with the cold bath of inverse temperature $\beta_c$ for a time $\tau_c$. The exchanged heat between the system and the cold bath is given by
\begin{align} \label{Eq: Cold isochore}
    Q_c =  \sum_{n=1}^{3} E_n(B) (P_n (\beta_c) - P_n (\beta_h)) .
\end{align}
At point $D$, the system reaches to a thermal state defined by inverse temperature $\beta_{c}$, and no work is done during this stage.
\par \textbf{Adiabatic Compression ($D\rightarrow A$)}: Similar to adiabatic expansion, the system is isolated from the environment during this stage. The eigenenergies are changed by varying either phase $\Phi$ or detuning $\delta$; the occupation probabilities remain the same during this stroke. The cycle closes with the adiabatic compression while the system Hamiltonian shifts slowly from $H(B)$ to $H(A)$ for a time $\tau_4$.
The net produced work during a cycle can be calculated by using Eqs.~(\ref{Eq: Hot isochore} and \ref{Eq: Cold isochore})
\begin{align} \label{Eq: Work1}
    W_{0} &= Q_h + Q_c \nonumber \\ 
    &= \sum_{n=1}^{3} [E_n(A) - E_n(B)](P_n (\beta_h) - P_n (\beta_c)).
\end{align}
The efficiency of a heat engine is defined as $\eta=W_{0}/Q_{h}$. Here we adopt the convention that the work done on (by) the system is positive, and heat flow out of (into) the system is negative.

\section{ Results} \label{sec: results}
We look for the signatures in the work output of the Otto cycle to distinguish left- and right-handed enantiomers. The system's Hamiltonian (Eq.(\ref{eq:01Ham})) is a function of the overall phase $\Phi$ and detuning $\delta$.  As a result, the adiabatic stages of the cycle can be implemented either by varying the phase or detuning and keeping the other constant during the stroke. In this section, we examine both strategies for implementing the Otto cycle; and evaluate the output work to distinguish enantiomers. First, we investigate the Otto cycle in constant detuning $\delta$ regime while the overall phase employs as the control parameter in Sec.~\ref{sec: ConstantDetuning}, and then in Sec.~\ref{sec:ConstantPhase} examine the Otto cycle when $\delta$ is the control parameter, and overall phase kept constant. 
	\begin{figure}[t]
		\centering
		\subfloat[]{
			\includegraphics[width = 0.9 \linewidth]{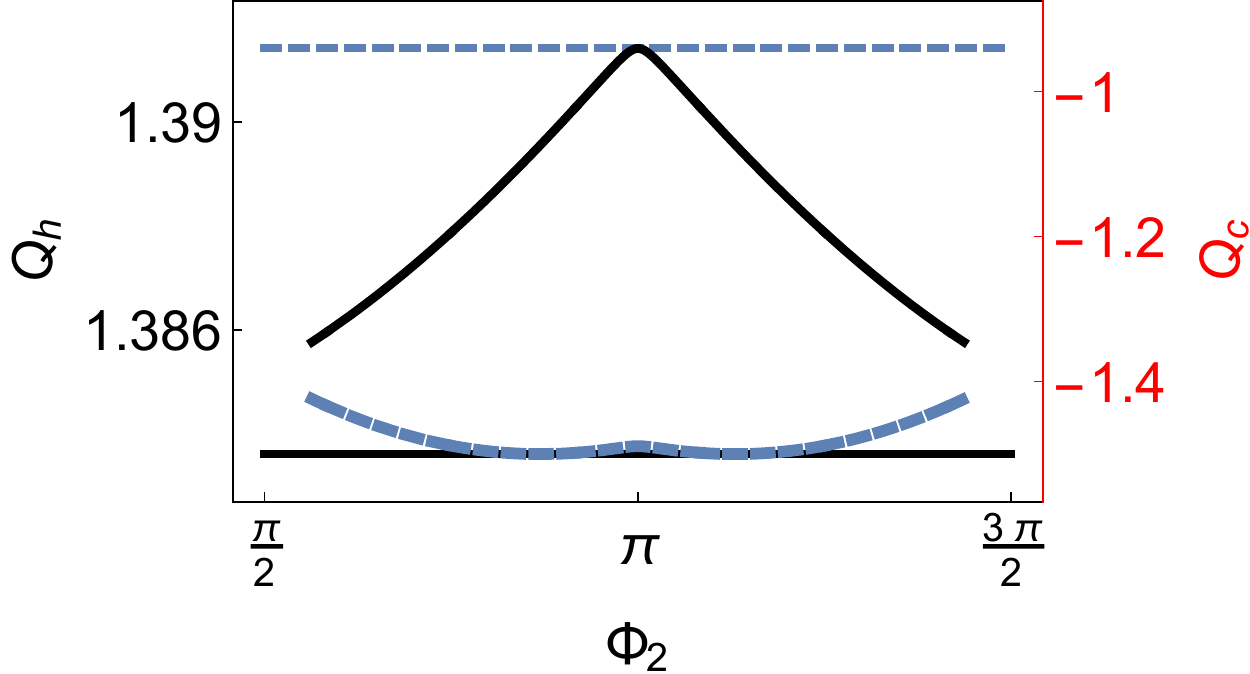}}\\
		\subfloat[]{
			\includegraphics[width = 0.75 \linewidth]{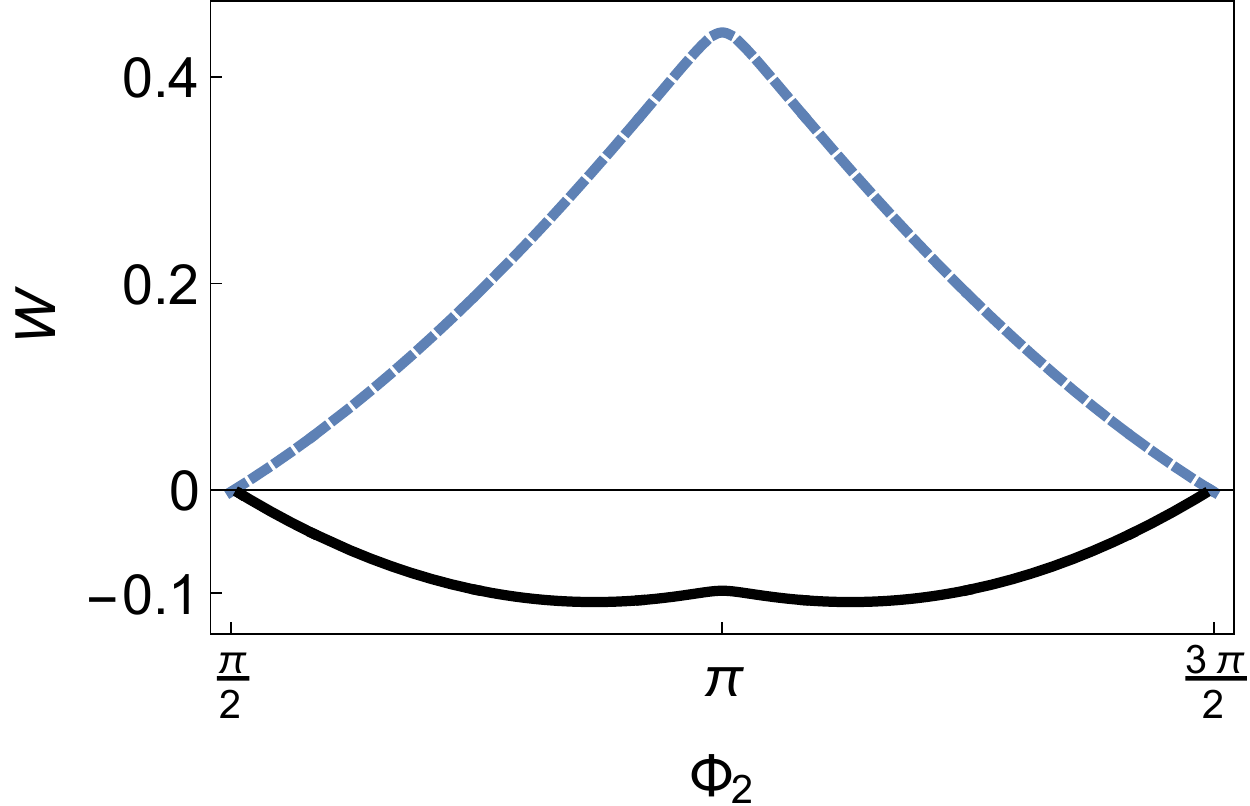}}
		\caption{(Color online)  (a) Heat exchange $Q_{h}$ with the hot bath and $Q_{c}$ (scaled by $E_0$) with the cold bath as a function of the overall phase $\Phi$ of the optical drives with contant $\delta = 0.1/\tau_0$. The solid and dashed lines are for the right- and left-handed enantiomers. In addition, upper and lower horizontal lines represent heat $Q_{h}$ injected into the system, and heat rejected $Q_{c}$ into the cold bath is denoted by curved lines. (b) Work output $W$ as a function of the phase $\Phi$ of the optical drives in the quantum Otto cycle operating between the hot and cold baths of inverse temperatures $\beta_{h}=0.01$, and $\beta_{c}=1$, respectively. The solid and dashed lines are for the left- and right-handed enantiomers, respectively.}\label{fig:fig5}
	\end{figure}

\subsection{Control parameter $\Phi$} \label{sec: ConstantDetuning}
We adhere to the sign conventions of absorbed (rejected) heat is positive (negative), and work extracted (consumed) is negative. Accordingly, the thermal machines' operations can be categorized into four distinct regimes~\cite{thermalMachines001}
\small
\begin{align}
 &Engine: & W \leq 0, Q_h \geq 0, Q_c \leq 0,\nonumber \\
 &Refrigerator: & W \geq 0, Q_h \leq 0, Q_c \geq 0, \nonumber \\
 &Heater: & W \geq 0, Q_h \leq 0, Q_c \leq 0, \nonumber \\ 
 &Thermal~accelerator: & W \geq 0, Q_h \geq 0, Q_c \leq 0. \nonumber
\end{align}
\normalsize
We consider the set of system parameters for which there is no level crossing between the energy levels during the Otto cycle (see Fig.~\ref{fig:fig2}). Accordingly, we consider the phase $\Phi_1 = \pi/2$ at the start of the adiabatic expansion and change it to  $\Phi_2 = 3\pi/2$ at the end of this stage. Similarly, the phase is changed from  $\Phi_2$ to $\Phi_1$  in the adiabatic compression stage. During the adiabatic stages detuning $\delta = 0.1/\tau_0$ is kept constant. The heat exchanged with the thermal baths and extracted work for left- and right-handed systems are shown in Fig.~\ref{fig:fig5}. For both left- and right-handed enantiomers, the heat injected $Q_{h}$ from the hot bath into the system is positive, and heat rejected $Q_{c}$ into the cold bath is negative. However, the work distribution reveals that work is extracted from the left-handed system ($W<0$) but in the right-handed system, the work is done on the system ($W>0$). Accordingly, the left- and right-handed enantiomers operate as a heat engine and thermal accelerator, respectively. Therefore, the thermodynamics response of the enantiomers are not the same, which can be used as a probe to distinguish these enantiomers. 

\begin{figure}[t]
    \centering
     \includegraphics[width = 0.8 \linewidth]{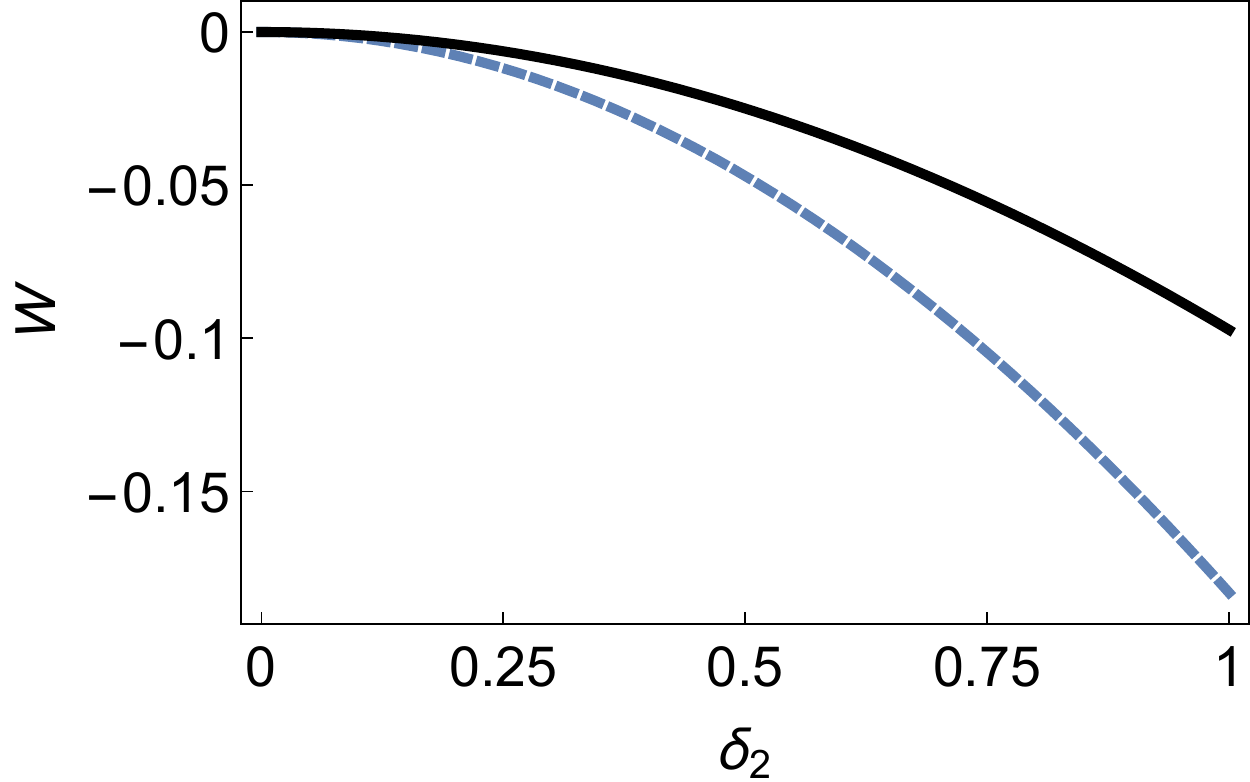}
    \caption{(Color online) Work output $W$ as a function of the detuning $\delta_2$ (scaled by $\tau_0$) supposing the optical drive fixed initially at resonance, $\delta_1 = 0$ and overall phase kept constant at $\Phi = 0$. The quantum Otto cycle operates between the hot and cold baths of inverse temperatures $\beta_{h}=0.01$, and $\beta_{c}=1$, respectively. The solid and dashed lines are for the left- and right-handed enantiomers.}
    \label{fig:fig6}
\end{figure}

\subsection{Control parameter $\delta$}\label{sec:ConstantPhase}
 	It can be of practical importance to check the thermal behaviors of chiral three-level systems in the Otto cycle using $\delta$ as the control parameter with a constant phase $\Phi$. The work output, in this case, is shown in Fig.~\ref{fig:fig6}, which shows that both the left- and right-handed enantiomers operate as heat engines. However, the extracted work in both cases can be significantly different.
 	The extracted work differs between left- and right-handed systems based on the constant $\Phi$ selection. For instance, if we keep overall phase $\Phi = \pi/2$ during the Otto cycle, the obtained extracted work is the same for both enantiomers, but it is significantly different for $\Phi = \pi$ where $W_{left} = - 0.18~E_0$ and $W_{right} = - 0.097~E_0$. \\
 \par
\begin{figure}[ht]
    \centering
     \includegraphics[width = 0.8 \linewidth]{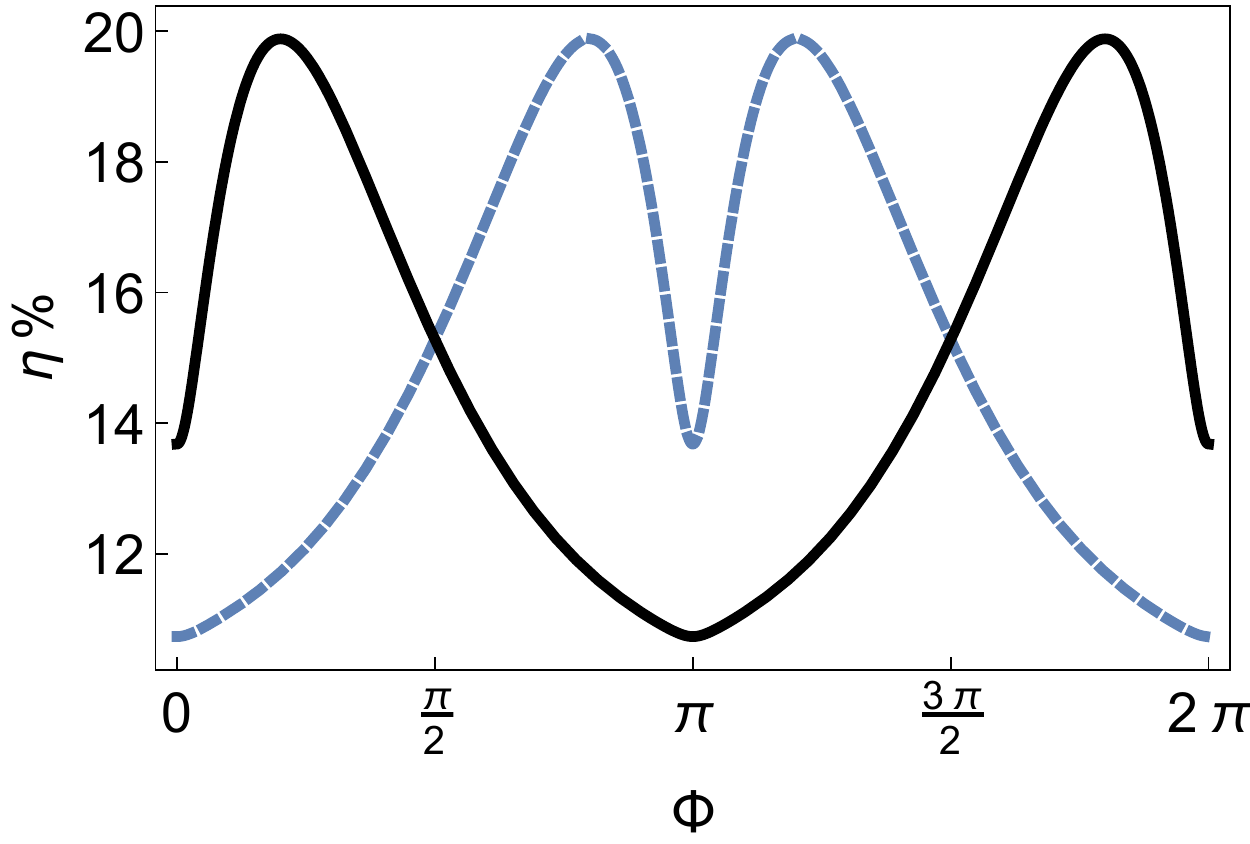}
    \caption{(Color online) The engine's efficiency $\eta (\%)$ as a function of $\Phi$, while detuning as a control parameter modifies from $\delta = 0$ to $\delta = 1/\tau_0$, in the quantum Otto cycle operating between the hot and cold baths of inverse temperatures $\beta_{h}=0.01/\tau_0$, and $\beta_{c}=1/\tau_0$, respectively. The solid and dashed lines are for the left- and right-handed Otto engines.}
    \label{fig:fig7}
\end{figure}

 The efficiency of the Otto engine is given by
 	\begin{align}
 	    \eta (\%) = \frac{|W|}{Q_h} \times 100.
 	\end{align}
 	To study the efficiency, we consider the control parameter $\delta$ modifying from $\delta_1 = 0$ to $\delta_2 = 1/\tau_0$. We define the efficiency as a function of the overall phase $\Phi$, which is constant during the cycle. The engine's efficiency corresponds to $\Phi$ is plotted in Fig.~\ref{fig:fig7} where the solid black and dashed blue curves denote the left- and right-handed system's efficiency, respectively. According to Fig.~\ref{fig:fig7}, the magnitude of maximum efficiency for both cases is the same, $\eta \approx 20 \%$. The curves intersect, so both are equally efficient in a particular $\Phi$. However, their efficiency is distinct at other $\Phi$ values. For example, the left-handed Otto engine's maximum efficiency is obtained at $\Phi \approx \pi/5$, and $6\pi/5$ while it is maximum at $\Phi \approx \pi \pm 0.7$ for the right-handed engine. Also, $\eta_{\text{left}} = \eta_{\text{right}}$ at $\Phi = \pi/2$, and $3\pi/2$ which comes from their identical energy spectrum in these points (Fig.~\ref{fig:fig2}). Therefore, one enantiomer may be more favorable for some energy processes because of their distinguishing efficiency, even if they can perform the same energetic function (such as operating a heat engine).

\section{Conclusion}\label{sec:conclusion}
We have investigated the possibility of exploiting thermodynamic processes for enantiomers detection. In particular, we considered a chiral molecule described by a cyclic three-level system coupled to three external classical optical fields. We found that depending on the phase and detuning of the optical fields, the left- and right-handed cyclic three-level system exhibited different thermal functions when subjected to an Otto cycle. Specifically, suppose the adiabatic strokes in the cycle are implemented by changing the overall phase of three optical drives. In that case, the left- and right-handed enantiomer operate as a quantum heat engine and thermal accelerator. Suppose the adiabatic strokes in the cycle are implemented by changing the detuning $\delta$ and keeping phase $\Phi$ constant during the cycle, which is more practical from the experimental point of view. In that case, both the left- and right-handed enantiomers operate as heat engines. However, the enantiomers still can be distinguished by the output work distribution and their efficiency, as quantitatively. In addition to the practical distinction of enantiomers, our results fundamentally suggest that different enantiomers may be associated with different thermodynamic (energetic) functions in chemical or biological processes, such as work harvesting. Besides, even if they are capable of performing the same energetic function, their efficiencies may differ so that one enantiomer may be more favorable than the other. One of the crucial steps in implementing the Otto cycle with a three-level system with cyclic optical transitions is the thermalization of the system. In general, in the presence of external laser drives, a system coupled to a thermal bath may not reach a thermal state. However, by numerically solving the master equation for the isochoric stages of the cycle, we showed that a thermal state could approximately give the system's state.

	Enantiomer detection is a formidable task, even for the case of two enantiomers, because of their identical energy spectra \cite{QUACK1986147,cphc.201500801,chiral001}. Our analysis revealed that thermodynamic processes could be exploited for enantiomer detection. Our method of enantiomer detection via work distribution can be an alternative to more widely employed schemes for discrimination between the enantiomers \cite{Intro-select01,Intro-select07}. 
\begin{acknowledgments}
This work was supported by the Scientific and Technological Research Council of
Turkey (TUBITAK) with project number 120F089. \\
The authors thank Dr.  Onur Pusuluk for the helpful discussions.
\end{acknowledgments}

\bibliography{ChiralHE}

\providecommand{\noopsort}[1]{}\providecommand{\singleletter}[1]{#1}%
\begin{thebibliography}{59}%
\makeatletter
\providecommand \@ifxundefined [1]{%
 \@ifx{#1\undefined}
}%
\providecommand \@ifnum [1]{%
 \ifnum #1\expandafter \@firstoftwo
 \else \expandafter \@secondoftwo
 \fi
}%
\providecommand \@ifx [1]{%
 \ifx #1\expandafter \@firstoftwo
 \else \expandafter \@secondoftwo
 \fi
}%
\providecommand \natexlab [1]{#1}%
\providecommand \enquote  [1]{``#1''}%
\providecommand \bibnamefont  [1]{#1}%
\providecommand \bibfnamefont [1]{#1}%
\providecommand \citenamefont [1]{#1}%
\providecommand \href@noop [0]{\@secondoftwo}%
\providecommand \href [0]{\begingroup \@sanitize@url \@href}%
\providecommand \@href[1]{\@@startlink{#1}\@@href}%
\providecommand \@@href[1]{\endgroup#1\@@endlink}%
\providecommand \@sanitize@url [0]{\catcode `\\12\catcode `\$12\catcode
  `\&12\catcode `\#12\catcode `\^12\catcode `\_12\catcode `\%12\relax}%
\providecommand \@@startlink[1]{}%
\providecommand \@@endlink[0]{}%
\providecommand \url  [0]{\begingroup\@sanitize@url \@url }%
\providecommand \@url [1]{\endgroup\@href {#1}{\urlprefix }}%
\providecommand \urlprefix  [0]{URL }%
\providecommand \Eprint [0]{\href }%
\providecommand \doibase [0]{http://dx.doi.org/}%
\providecommand \selectlanguage [0]{\@gobble}%
\providecommand \bibinfo  [0]{\@secondoftwo}%
\providecommand \bibfield  [0]{\@secondoftwo}%
\providecommand \translation [1]{[#1]}%
\providecommand \BibitemOpen [0]{}%
\providecommand \bibitemStop [0]{}%
\providecommand \bibitemNoStop [0]{.\EOS\space}%
\providecommand \EOS [0]{\spacefactor3000\relax}%
\providecommand \BibitemShut  [1]{\csname bibitem#1\endcsname}%
\let\auto@bib@innerbib\@empty
\bibitem [{\citenamefont {J.}\ \emph {et~al.}(2007)\citenamefont {J.},
  \citenamefont {Micheal},\ and\ \citenamefont {Mahler}}]{QTBook01}%
  \BibitemOpen
  \bibfield  {author} {\bibinfo {author} {\bibfnamefont {Gemmer}\ \bibnamefont
  {J.}}, \bibinfo {author} {\bibfnamefont {M.}~\bibnamefont {Micheal}}, \ and\
  \bibinfo {author} {\bibfnamefont {G.}~\bibnamefont {Mahler}},\ }\href
  {\doibase 10.1007/978-3-540-70510-9} {\emph {\bibinfo {title} {{Quantum
  Thermodynamics}}}}\ (\bibinfo  {publisher} {Oxford University Press},\
  \bibinfo {year} {2007})\BibitemShut {NoStop}%
\bibitem [{\citenamefont {Kosloff}(2013)}]{e15062100}%
  \BibitemOpen
  \bibfield  {author} {\bibinfo {author} {\bibfnamefont {Ronnie}\ \bibnamefont
  {Kosloff}},\ }\bibfield  {title} {\enquote {\bibinfo {title} {Quantum
  thermodynamics: A dynamical viewpoint},}\ }\href {\doibase 10.3390/e15062100}
  {\bibfield  {journal} {\bibinfo  {journal} {Entropy}\ }\textbf {\bibinfo
  {volume} {15}},\ \bibinfo {pages} {2100--2128} (\bibinfo {year}
  {2013})}\BibitemShut {NoStop}%
\bibitem [{\citenamefont {Vinjanampathy}\ and\ \citenamefont
  {Anders}(2016)}]{00107514}%
  \BibitemOpen
  \bibfield  {author} {\bibinfo {author} {\bibfnamefont {Sai}\ \bibnamefont
  {Vinjanampathy}}\ and\ \bibinfo {author} {\bibfnamefont {Janet}\ \bibnamefont
  {Anders}},\ }\bibfield  {title} {\enquote {\bibinfo {title} {Quantum
  thermodynamics},}\ }\href {\doibase 10.1080/00107514.2016.1201896} {\bibfield
   {journal} {\bibinfo  {journal} {Contemp. Phys.}\ }\textbf {\bibinfo {volume}
  {57}},\ \bibinfo {pages} {545--579} (\bibinfo {year} {2016})}\BibitemShut
  {NoStop}%
\bibitem [{\citenamefont {Goold}\ \emph {et~al.}(2016)\citenamefont {Goold},
  \citenamefont {Huber}, \citenamefont {Riera}, \citenamefont {del Rio},\ and\
  \citenamefont {Skrzypczyk}}]{Goold_2016}%
  \BibitemOpen
  \bibfield  {author} {\bibinfo {author} {\bibfnamefont {John}\ \bibnamefont
  {Goold}}, \bibinfo {author} {\bibfnamefont {Marcus}\ \bibnamefont {Huber}},
  \bibinfo {author} {\bibfnamefont {Arnau}\ \bibnamefont {Riera}}, \bibinfo
  {author} {\bibfnamefont {Lídia}\ \bibnamefont {del Rio}}, \ and\ \bibinfo
  {author} {\bibfnamefont {Paul}\ \bibnamefont {Skrzypczyk}},\ }\bibfield
  {title} {\enquote {\bibinfo {title} {The role of quantum information in
  thermodynamics—a topical review},}\ }\href {\doibase
  10.1088/1751-8113/49/14/143001} {\bibfield  {journal} {\bibinfo  {journal}
  {J. Phys. A: Math. Theor.}\ }\textbf {\bibinfo {volume} {49}},\ \bibinfo
  {pages} {143001} (\bibinfo {year} {2016})}\BibitemShut {NoStop}%
\bibitem [{\citenamefont {Deffner}\ and\ \citenamefont
  {Campbell}(2019)}]{QTBook02}%
  \BibitemOpen
  \bibfield  {author} {\bibinfo {author} {\bibfnamefont {Sebastian}\
  \bibnamefont {Deffner}}\ and\ \bibinfo {author} {\bibfnamefont {Steve}\
  \bibnamefont {Campbell}},\ }\href {\doibase 10.1088/2053-2571/ab21c6} {\emph
  {\bibinfo {title} {Quantum Thermodynamics}}},\ 2053-2571\ (\bibinfo
  {publisher} {Morgan Claypool Publishers},\ \bibinfo {year}
  {2019})\BibitemShut {NoStop}%
\bibitem [{\citenamefont {Tuncer}\ \emph {et~al.}(2020)\citenamefont {Tuncer},
  ,\ and\ \citenamefont {M\"ustecapl\ifmmode \imath \else \i
  \fi{}o\ifmmode~\breve{g}\else \u{g}\fi{}lu}}]{AsliReview}%
  \BibitemOpen
  \bibfield  {author} {\bibinfo {author} {\bibfnamefont {A.}~\bibnamefont
  {Tuncer}}, , \ and\ \bibinfo {author} {\bibfnamefont {\"Ozg\"ur~E.}\
  \bibnamefont {M\"ustecapl\ifmmode \imath \else \i
  \fi{}o\ifmmode~\breve{g}\else \u{g}\fi{}lu}},\ }\bibfield  {title} {\enquote
  {\bibinfo {title} {Quantum thermodynamics and quantum coherence engines},}\
  }\href {\doibase 10.3906/fiz-2009-12} {\bibfield  {journal} {\bibinfo
  {journal} {Turk. J. Phys.}\ }\textbf {\bibinfo {volume} {44}},\ \bibinfo
  {pages} {404} (\bibinfo {year} {2020})}\BibitemShut {NoStop}%
\bibitem [{\citenamefont {Myers}\ \emph {et~al.}(2022)\citenamefont {Myers},
  \citenamefont {Abah},\ and\ \citenamefont {Deffner}}]{5.0083192}%
  \BibitemOpen
  \bibfield  {author} {\bibinfo {author} {\bibfnamefont {Nathan~M.}\
  \bibnamefont {Myers}}, \bibinfo {author} {\bibfnamefont {Obinna}\
  \bibnamefont {Abah}}, \ and\ \bibinfo {author} {\bibfnamefont {Sebastian}\
  \bibnamefont {Deffner}},\ }\bibfield  {title} {\enquote {\bibinfo {title}
  {Quantum thermodynamic devices: From theoretical proposals to experimental
  reality},}\ }\href {\doibase 10.1116/5.0083192} {\bibfield  {journal}
  {\bibinfo  {journal} {AVS Quantum Science}\ }\textbf {\bibinfo {volume}
  {4}},\ \bibinfo {pages} {027101} (\bibinfo {year} {2022})}\BibitemShut
  {NoStop}%
\bibitem [{\citenamefont {Kosloff}\ and\ \citenamefont
  {Levy}(2014)}]{Intro-QHE01}%
  \BibitemOpen
  \bibfield  {author} {\bibinfo {author} {\bibfnamefont {Ronnie}\ \bibnamefont
  {Kosloff}}\ and\ \bibinfo {author} {\bibfnamefont {Amikam}\ \bibnamefont
  {Levy}},\ }\bibfield  {title} {\enquote {\bibinfo {title} {Quantum heat
  engines and refrigerators: Continuous devices},}\ }\href {\doibase
  10.1146/annurev-physchem-040513-103724} {\bibfield  {journal} {\bibinfo
  {journal} {Annu. Rev. Phys. Chem.}\ }\textbf {\bibinfo {volume} {65}},\
  \bibinfo {pages} {365--393} (\bibinfo {year} {2014})}\BibitemShut {NoStop}%
\bibitem [{\citenamefont {Kosloff}\ and\ \citenamefont
  {Rezek}(2017)}]{Intro-QOE01}%
  \BibitemOpen
  \bibfield  {author} {\bibinfo {author} {\bibfnamefont {Ronnie}\ \bibnamefont
  {Kosloff}}\ and\ \bibinfo {author} {\bibfnamefont {Yair}\ \bibnamefont
  {Rezek}},\ }\bibfield  {title} {\enquote {\bibinfo {title} {The quantum
  harmonic otto cycle},}\ }\href {\doibase 10.3390/e19040136} {\bibfield
  {journal} {\bibinfo  {journal} {Entropy}\ }\textbf {\bibinfo {volume} {19}}
  (\bibinfo {year} {2017}),\ 10.3390/e19040136}\BibitemShut {NoStop}%
\bibitem [{\citenamefont {Izadyari}\ \emph {et~al.}(2022)\citenamefont
  {Izadyari}, \citenamefont {Öncü}, \citenamefont {Durak},\ and\
  \citenamefont {Mustecaplıoglu}}]{Intro-QOE02}%
  \BibitemOpen
  \bibfield  {author} {\bibinfo {author} {\bibfnamefont {M.}~\bibnamefont
  {Izadyari}}, \bibinfo {author} {\bibfnamefont {M.}~\bibnamefont {Öncü}},
  \bibinfo {author} {\bibfnamefont {K.}~\bibnamefont {Durak}}, \ and\ \bibinfo
  {author} {\bibfnamefont {O.~E.}\ \bibnamefont {Mustecaplıoglu}},\ }\bibfield
   {title} {\enquote {\bibinfo {title} {Quantum signatures in quadratic
  optomechanical heat engine with an atom in a tapered trap},}\ }\href@noop {}
  {\bibfield  {journal} {\bibinfo  {journal} {arXiv:2111.12803v3}\ } (\bibinfo
  {year} {2022})}\BibitemShut {NoStop}%
\bibitem [{\citenamefont {Altintas}\ and\ \citenamefont {M\"ustecapl\ifmmode
  \imath \else \i \fi{}o\ifmmode~\breve{g}\else
  \u{g}\fi{}lu}(2015)}]{Intro-QOE03}%
  \BibitemOpen
  \bibfield  {author} {\bibinfo {author} {\bibfnamefont {Ferdi}\ \bibnamefont
  {Altintas}}\ and\ \bibinfo {author} {\bibfnamefont {\"Ozg\"ur~E.}\
  \bibnamefont {M\"ustecapl\ifmmode \imath \else \i
  \fi{}o\ifmmode~\breve{g}\else \u{g}\fi{}lu}},\ }\bibfield  {title} {\enquote
  {\bibinfo {title} {General formalism of local thermodynamics with an example:
  Quantum otto engine with a spin-$1/2$ coupled to an arbitrary spin},}\ }\href
  {\doibase 10.1103/PhysRevE.92.022142} {\bibfield  {journal} {\bibinfo
  {journal} {Phys. Rev. E}\ }\textbf {\bibinfo {volume} {92}},\ \bibinfo
  {pages} {022142} (\bibinfo {year} {2015})}\BibitemShut {NoStop}%
\bibitem [{\citenamefont {Singh}\ \emph {et~al.}(2022)\citenamefont {Singh},
  \citenamefont {Singh}, \citenamefont {Abah},\ and\ \citenamefont
  {M\"ustecapl\ifmmode \imath \else \i \fi{}o\ifmmode~\breve{g}\else
  \u{g}\fi{}lu}}]{Intro-QOE04}%
  \BibitemOpen
  \bibfield  {author} {\bibinfo {author} {\bibfnamefont {Varinder}\
  \bibnamefont {Singh}}, \bibinfo {author} {\bibfnamefont {Satnam}\
  \bibnamefont {Singh}}, \bibinfo {author} {\bibfnamefont {Obinna}\
  \bibnamefont {Abah}}, \ and\ \bibinfo {author} {\bibfnamefont {\"Ozg\"ur~E.}\
  \bibnamefont {M\"ustecapl\ifmmode \imath \else \i
  \fi{}o\ifmmode~\breve{g}\else \u{g}\fi{}lu}},\ }\bibfield  {title} {\enquote
  {\bibinfo {title} {Unified trade-off optimization of quantum harmonic otto
  engine and refrigerator},}\ }\href {\doibase 10.1103/PhysRevE.106.024137}
  {\bibfield  {journal} {\bibinfo  {journal} {Phys. Rev. E}\ }\textbf {\bibinfo
  {volume} {106}},\ \bibinfo {pages} {024137} (\bibinfo {year}
  {2022})}\BibitemShut {NoStop}%
\bibitem [{\citenamefont {Scovil}\ and\ \citenamefont
  {Schulz-DuBois}(1959)}]{Intro-QHE02}%
  \BibitemOpen
  \bibfield  {author} {\bibinfo {author} {\bibfnamefont {H.~E.~D.}\
  \bibnamefont {Scovil}}\ and\ \bibinfo {author} {\bibfnamefont {E.~O.}\
  \bibnamefont {Schulz-DuBois}},\ }\bibfield  {title} {\enquote {\bibinfo
  {title} {Three-level masers as heat engines},}\ }\href {\doibase
  10.1103/PhysRevLett.2.262} {\bibfield  {journal} {\bibinfo  {journal} {Phys.
  Rev. Lett.}\ }\textbf {\bibinfo {volume} {2}},\ \bibinfo {pages} {262--263}
  (\bibinfo {year} {1959})}\BibitemShut {NoStop}%
\bibitem [{\citenamefont {Scully}\ \emph {et~al.}(2011)\citenamefont {Scully},
  \citenamefont {Chapin}, \citenamefont {Dorfman}, \citenamefont {Kim},\ and\
  \citenamefont {Svidzinsky}}]{Intro-QHE03}%
  \BibitemOpen
  \bibfield  {author} {\bibinfo {author} {\bibfnamefont {Marlan~O.}\
  \bibnamefont {Scully}}, \bibinfo {author} {\bibfnamefont {Kimberly~R.}\
  \bibnamefont {Chapin}}, \bibinfo {author} {\bibfnamefont {Konstantin~E.}\
  \bibnamefont {Dorfman}}, \bibinfo {author} {\bibfnamefont {Moochan~Barnabas}\
  \bibnamefont {Kim}}, \ and\ \bibinfo {author} {\bibfnamefont {Anatoly}\
  \bibnamefont {Svidzinsky}},\ }\bibfield  {title} {\enquote {\bibinfo {title}
  {Quantum heat engine power can be increased by noise-induced coherence},}\
  }\href {\doibase 10.1073/pnas.1110234108} {\bibfield  {journal} {\bibinfo
  {journal} {Proc. Natl Acad. Sci.}\ }\textbf {\bibinfo {volume} {108}},\
  \bibinfo {pages} {15097--15100} (\bibinfo {year} {2011})}\BibitemShut
  {NoStop}%
\bibitem [{\citenamefont {Thomas}\ and\ \citenamefont
  {Johal}(2011)}]{Intro-QHE04}%
  \BibitemOpen
  \bibfield  {author} {\bibinfo {author} {\bibfnamefont {George}\ \bibnamefont
  {Thomas}}\ and\ \bibinfo {author} {\bibfnamefont {Ramandeep~S.}\ \bibnamefont
  {Johal}},\ }\bibfield  {title} {\enquote {\bibinfo {title} {Coupled quantum
  otto cycle},}\ }\href {\doibase 10.1103/PhysRevE.83.031135} {\bibfield
  {journal} {\bibinfo  {journal} {Phys. Rev. E}\ }\textbf {\bibinfo {volume}
  {83}},\ \bibinfo {pages} {031135} (\bibinfo {year} {2011})}\BibitemShut
  {NoStop}%
\bibitem [{\citenamefont {Zhang}\ \emph {et~al.}(2014)\citenamefont {Zhang},
  \citenamefont {Bariani},\ and\ \citenamefont {Meystre}}]{Intro-QHE05}%
  \BibitemOpen
  \bibfield  {author} {\bibinfo {author} {\bibfnamefont {Keye}\ \bibnamefont
  {Zhang}}, \bibinfo {author} {\bibfnamefont {Francesco}\ \bibnamefont
  {Bariani}}, \ and\ \bibinfo {author} {\bibfnamefont {Pierre}\ \bibnamefont
  {Meystre}},\ }\bibfield  {title} {\enquote {\bibinfo {title} {Quantum
  optomechanical heat engine},}\ }\href {\doibase
  10.1103/PhysRevLett.112.150602} {\bibfield  {journal} {\bibinfo  {journal}
  {Phys. Rev. Lett.}\ }\textbf {\bibinfo {volume} {112}},\ \bibinfo {pages}
  {150602} (\bibinfo {year} {2014})}\BibitemShut {NoStop}%
\bibitem [{\citenamefont {Insinga}\ \emph {et~al.}(2016)\citenamefont
  {Insinga}, \citenamefont {Andresen},\ and\ \citenamefont
  {Salamon}}]{Intro-QHE06}%
  \BibitemOpen
  \bibfield  {author} {\bibinfo {author} {\bibfnamefont {Andrea}\ \bibnamefont
  {Insinga}}, \bibinfo {author} {\bibfnamefont {Bjarne}\ \bibnamefont
  {Andresen}}, \ and\ \bibinfo {author} {\bibfnamefont {Peter}\ \bibnamefont
  {Salamon}},\ }\bibfield  {title} {\enquote {\bibinfo {title} {Thermodynamical
  analysis of a quantum heat engine based on harmonic oscillators},}\ }\href
  {\doibase 10.1103/PhysRevE.94.012119} {\bibfield  {journal} {\bibinfo
  {journal} {Phys. Rev. E}\ }\textbf {\bibinfo {volume} {94}},\ \bibinfo
  {pages} {012119} (\bibinfo {year} {2016})}\BibitemShut {NoStop}%
\bibitem [{\citenamefont {Naseem}\ and\ \citenamefont {\"{O}zg\"{u}r
  E.~M\"{u}stecaplio\u{g}lu}(2019)}]{Intro-QHE07}%
  \BibitemOpen
  \bibfield  {author} {\bibinfo {author} {\bibfnamefont {M.~Tahir}\
  \bibnamefont {Naseem}}\ and\ \bibinfo {author} {\bibnamefont {\"{O}zg\"{u}r
  E.~M\"{u}stecaplio\u{g}lu}},\ }\bibfield  {title} {\enquote {\bibinfo {title}
  {Quantum heat engine with a quadratically coupled optomechanical system},}\
  }\href {\doibase 10.1364/JOSAB.36.003000} {\bibfield  {journal} {\bibinfo
  {journal} {J. Opt. Soc. Am. B}\ }\textbf {\bibinfo {volume} {36}},\ \bibinfo
  {pages} {3000--3008} (\bibinfo {year} {2019})}\BibitemShut {NoStop}%
\bibitem [{\citenamefont {Peterson}\ \emph {et~al.}(2019)\citenamefont
  {Peterson}, \citenamefont {Batalh\~ao}, \citenamefont {Herrera},
  \citenamefont {Souza}, \citenamefont {Sarthour}, \citenamefont {Oliveira},\
  and\ \citenamefont {Serra}}]{Intro-QHE08}%
  \BibitemOpen
  \bibfield  {author} {\bibinfo {author} {\bibfnamefont {John P.~S.}\
  \bibnamefont {Peterson}}, \bibinfo {author} {\bibfnamefont {Tiago~B.}\
  \bibnamefont {Batalh\~ao}}, \bibinfo {author} {\bibfnamefont {Marcela}\
  \bibnamefont {Herrera}}, \bibinfo {author} {\bibfnamefont {Alexandre~M.}\
  \bibnamefont {Souza}}, \bibinfo {author} {\bibfnamefont {Roberto~S.}\
  \bibnamefont {Sarthour}}, \bibinfo {author} {\bibfnamefont {Ivan~S.}\
  \bibnamefont {Oliveira}}, \ and\ \bibinfo {author} {\bibfnamefont
  {Roberto~M.}\ \bibnamefont {Serra}},\ }\bibfield  {title} {\enquote {\bibinfo
  {title} {Experimental characterization of a spin quantum heat engine},}\
  }\href {\doibase 10.1103/PhysRevLett.123.240601} {\bibfield  {journal}
  {\bibinfo  {journal} {Phys. Rev. Lett.}\ }\textbf {\bibinfo {volume} {123}},\
  \bibinfo {pages} {240601} (\bibinfo {year} {2019})}\BibitemShut {NoStop}%
\bibitem [{\citenamefont {Bouton}\ \emph {et~al.}(2021)\citenamefont {Bouton},
  \citenamefont {Nettersheim}, \citenamefont {Burgardt}, \citenamefont {Adam},
  \citenamefont {Lutz},\ and\ \citenamefont {Widera}}]{Bouton2021}%
  \BibitemOpen
  \bibfield  {author} {\bibinfo {author} {\bibfnamefont {Quentin}\ \bibnamefont
  {Bouton}}, \bibinfo {author} {\bibfnamefont {Jens}\ \bibnamefont
  {Nettersheim}}, \bibinfo {author} {\bibfnamefont {Sabrina}\ \bibnamefont
  {Burgardt}}, \bibinfo {author} {\bibfnamefont {Daniel}\ \bibnamefont {Adam}},
  \bibinfo {author} {\bibfnamefont {Eric}\ \bibnamefont {Lutz}}, \ and\
  \bibinfo {author} {\bibfnamefont {Artur}\ \bibnamefont {Widera}},\ }\bibfield
   {title} {\enquote {\bibinfo {title} {A quantum heat engine driven by atomic
  collisions},}\ }\href {\doibase 10.1038/s41467-021-22222-z} {\bibfield
  {journal} {\bibinfo  {journal} {Nat. Commun.}\ }\textbf {\bibinfo {volume}
  {12}},\ \bibinfo {pages} {2063} (\bibinfo {year} {2021})}\BibitemShut
  {NoStop}%
\bibitem [{\citenamefont {Zhang}\ \emph {et~al.}(2022)\citenamefont {Zhang},
  \citenamefont {Zhang}, \citenamefont {Ding}, \citenamefont {Li},
  \citenamefont {Bu}, \citenamefont {Wang}, \citenamefont {Yan}, \citenamefont
  {Su}, \citenamefont {Chen}, \citenamefont {Nori}, \citenamefont
  {{\"O}zdemir}, \citenamefont {Zhou}, \citenamefont {Jing},\ and\
  \citenamefont {Feng}}]{Zhang2022}%
  \BibitemOpen
  \bibfield  {author} {\bibinfo {author} {\bibfnamefont {J.-W.}\ \bibnamefont
  {Zhang}}, \bibinfo {author} {\bibfnamefont {J.-Q.}\ \bibnamefont {Zhang}},
  \bibinfo {author} {\bibfnamefont {G.-Y.}\ \bibnamefont {Ding}}, \bibinfo
  {author} {\bibfnamefont {J.-C.}\ \bibnamefont {Li}}, \bibinfo {author}
  {\bibfnamefont {J.-T.}\ \bibnamefont {Bu}}, \bibinfo {author} {\bibfnamefont
  {B.}~\bibnamefont {Wang}}, \bibinfo {author} {\bibfnamefont {L.-L.}\
  \bibnamefont {Yan}}, \bibinfo {author} {\bibfnamefont {S.-L.}\ \bibnamefont
  {Su}}, \bibinfo {author} {\bibfnamefont {L.}~\bibnamefont {Chen}}, \bibinfo
  {author} {\bibfnamefont {F.}~\bibnamefont {Nori}}, \bibinfo {author}
  {\bibfnamefont {{\c{S}}~K.}\ \bibnamefont {{\"O}zdemir}}, \bibinfo {author}
  {\bibfnamefont {F.}~\bibnamefont {Zhou}}, \bibinfo {author} {\bibfnamefont
  {H.}~\bibnamefont {Jing}}, \ and\ \bibinfo {author} {\bibfnamefont
  {M.}~\bibnamefont {Feng}},\ }\bibfield  {title} {\enquote {\bibinfo {title}
  {Dynamical control of quantum heat engines using exceptional points},}\
  }\href {\doibase 10.1038/s41467-022-33667-1} {\bibfield  {journal} {\bibinfo
  {journal} {Nat. Commun.}\ }\textbf {\bibinfo {volume} {13}},\ \bibinfo
  {pages} {6225} (\bibinfo {year} {2022})}\BibitemShut {NoStop}%
\bibitem [{\citenamefont {Fox}\ and\ \citenamefont
  {Whitesell}(1974)}]{ChiralBook01}%
  \BibitemOpen
  \bibfield  {author} {\bibinfo {author} {\bibfnamefont {M.~A.}\ \bibnamefont
  {Fox}}\ and\ \bibinfo {author} {\bibfnamefont {J.~K.}\ \bibnamefont
  {Whitesell}},\ }\href@noop {} {\emph {\bibinfo {title} {Organic Chemistry}}}\
  (\bibinfo  {publisher} {Jones and Bartlett Publishers},\ \bibinfo {year}
  {1974})\BibitemShut {NoStop}%
\bibitem [{\citenamefont {Woolley}(1976)}]{chiral0011}%
  \BibitemOpen
  \bibfield  {author} {\bibinfo {author} {\bibfnamefont {R.G.}\ \bibnamefont
  {Woolley}},\ }\bibfield  {title} {\enquote {\bibinfo {title} {Quantum theory
  and molecular structure},}\ }\href {\doibase 10.1080/00018737600101352}
  {\bibfield  {journal} {\bibinfo  {journal} {Adv. Phys.}\ }\textbf {\bibinfo
  {volume} {25}},\ \bibinfo {pages} {27--52} (\bibinfo {year}
  {1976})}\BibitemShut {NoStop}%
\bibitem [{\citenamefont {Saito}\ and\ \citenamefont
  {Hyuga}(2013)}]{Intro-bio02}%
  \BibitemOpen
  \bibfield  {author} {\bibinfo {author} {\bibfnamefont {Yukio}\ \bibnamefont
  {Saito}}\ and\ \bibinfo {author} {\bibfnamefont {Hiroyuki}\ \bibnamefont
  {Hyuga}},\ }\bibfield  {title} {\enquote {\bibinfo {title} {Colloquium:
  Homochirality: Symmetry breaking in systems driven far from equilibrium},}\
  }\href {\doibase 10.1103/RevModPhys.85.603} {\bibfield  {journal} {\bibinfo
  {journal} {Rev. Mod. Phys.}\ }\textbf {\bibinfo {volume} {85}},\ \bibinfo
  {pages} {603--621} (\bibinfo {year} {2013})}\BibitemShut {NoStop}%
\bibitem [{\citenamefont {Gal}(2012)}]{Intro-bio03}%
  \BibitemOpen
  \bibfield  {author} {\bibinfo {author} {\bibfnamefont {Joseph}\ \bibnamefont
  {Gal}},\ }\bibfield  {title} {\enquote {\bibinfo {title} {The discovery of
  stereoselectivity at biological receptors: Arnaldo piutti and the taste of
  the asparagine enantiomers—history and analysis on the 125th
  anniversary},}\ }\href {\doibase https://doi.org/10.1002/chir.22071}
  {\bibfield  {journal} {\bibinfo  {journal} {Chirality}\ }\textbf {\bibinfo
  {volume} {24}},\ \bibinfo {pages} {959--976} (\bibinfo {year}
  {2012})}\BibitemShut {NoStop}%
\bibitem [{\citenamefont {Ariëns}(1986)}]{Intro-bio04}%
  \BibitemOpen
  \bibfield  {author} {\bibinfo {author} {\bibfnamefont {Everardus~J.}\
  \bibnamefont {Ariëns}},\ }\bibfield  {title} {\enquote {\bibinfo {title}
  {Stereochemistry: A source of problems in medicinal chemistry},}\ }\href
  {\doibase https://doi.org/10.1002/med.2610060404} {\bibfield  {journal}
  {\bibinfo  {journal} {Med. Res. Rev.}\ }\textbf {\bibinfo {volume} {6}},\
  \bibinfo {pages} {451--466} (\bibinfo {year} {1986})}\BibitemShut {NoStop}%
\bibitem [{\citenamefont {Quack}\ and\ \citenamefont
  {Stohner}(2000)}]{Intro-phys03}%
  \BibitemOpen
  \bibfield  {author} {\bibinfo {author} {\bibfnamefont {Martin}\ \bibnamefont
  {Quack}}\ and\ \bibinfo {author} {\bibfnamefont {J\"urgen}\ \bibnamefont
  {Stohner}},\ }\bibfield  {title} {\enquote {\bibinfo {title} {Influence of
  parity violating weak nuclear potentials on vibrational and rotational
  frequencies in chiral molecules},}\ }\href {\doibase
  10.1103/PhysRevLett.84.3807} {\bibfield  {journal} {\bibinfo  {journal}
  {Phys. Rev. Lett.}\ }\textbf {\bibinfo {volume} {84}},\ \bibinfo {pages}
  {3807--3810} (\bibinfo {year} {2000})}\BibitemShut {NoStop}%
\bibitem [{\citenamefont {Hutt}\ and\ \citenamefont
  {Tan}(1996)}]{Intro-farm01}%
  \BibitemOpen
  \bibfield  {author} {\bibinfo {author} {\bibfnamefont {A.~J.}\ \bibnamefont
  {Hutt}}\ and\ \bibinfo {author} {\bibfnamefont {S.~C.}\ \bibnamefont {Tan}},\
  }\bibfield  {title} {\enquote {\bibinfo {title} {Drug chirality and its
  clinical significance},}\ }\href {\doibase 10.2165/00003495-199600525-00003}
  {\bibfield  {journal} {\bibinfo  {journal} {Drugs}\ }\textbf {\bibinfo
  {volume} {52}},\ \bibinfo {pages} {1--12} (\bibinfo {year}
  {1996})}\BibitemShut {NoStop}%
\bibitem [{\citenamefont {Chen}\ \emph {et~al.}(2022)\citenamefont {Chen},
  \citenamefont {Cheng}, \citenamefont {Ye},\ and\ \citenamefont
  {Li}}]{Intro-phys04}%
  \BibitemOpen
  \bibfield  {author} {\bibinfo {author} {\bibfnamefont {Yu-Yuan}\ \bibnamefont
  {Chen}}, \bibinfo {author} {\bibfnamefont {Jian-Jian}\ \bibnamefont {Cheng}},
  \bibinfo {author} {\bibfnamefont {Chong}\ \bibnamefont {Ye}}, \ and\ \bibinfo
  {author} {\bibfnamefont {Yong}\ \bibnamefont {Li}},\ }\bibfield  {title}
  {\enquote {\bibinfo {title} {Enantiodetection of cyclic three-level chiral
  molecules in a driven cavity},}\ }\href {\doibase
  10.1103/PhysRevResearch.4.013100} {\bibfield  {journal} {\bibinfo  {journal}
  {Phys. Rev. Research}\ }\textbf {\bibinfo {volume} {4}},\ \bibinfo {pages}
  {013100} (\bibinfo {year} {2022})}\BibitemShut {NoStop}%
\bibitem [{\citenamefont {Ye}\ \emph {et~al.}(2021)\citenamefont {Ye},
  \citenamefont {Sun},\ and\ \citenamefont {Zhang}}]{Intro-select05}%
  \BibitemOpen
  \bibfield  {author} {\bibinfo {author} {\bibfnamefont {Chong}\ \bibnamefont
  {Ye}}, \bibinfo {author} {\bibfnamefont {Yifan}\ \bibnamefont {Sun}}, \ and\
  \bibinfo {author} {\bibfnamefont {Xiangdong}\ \bibnamefont {Zhang}},\
  }\bibfield  {title} {\enquote {\bibinfo {title} {Entanglement-assisted
  quantum chiral spectroscopy},}\ }\href {\doibase 10.1021/acs.jpclett.1c02196}
  {\bibfield  {journal} {\bibinfo  {journal} {Phys. Chem. Lett.}\ }\textbf
  {\bibinfo {volume} {12}},\ \bibinfo {pages} {8591--8597} (\bibinfo {year}
  {2021})}\BibitemShut {NoStop}%
\bibitem [{\citenamefont {Kondru}\ \emph {et~al.}(1998)\citenamefont {Kondru},
  \citenamefont {Wipf},\ and\ \citenamefont {Beratan}}]{Intro-select01}%
  \BibitemOpen
  \bibfield  {author} {\bibinfo {author} {\bibfnamefont {Rama~K.}\ \bibnamefont
  {Kondru}}, \bibinfo {author} {\bibfnamefont {Peter}\ \bibnamefont {Wipf}}, \
  and\ \bibinfo {author} {\bibfnamefont {David~N.}\ \bibnamefont {Beratan}},\
  }\bibfield  {title} {\enquote {\bibinfo {title} {Atomic contributions to the
  optical rotation angle as a quantitative probe of molecular chirality},}\
  }\href {\doibase 10.1126/science.282.5397.2247} {\bibfield  {journal}
  {\bibinfo  {journal} {Science}\ }\textbf {\bibinfo {volume} {282}},\ \bibinfo
  {pages} {2247--2250} (\bibinfo {year} {1998})}\BibitemShut {NoStop}%
\bibitem [{\citenamefont {Bielski}\ and\ \citenamefont
  {Tencer}(2005)}]{Intro-select02}%
  \BibitemOpen
  \bibfield  {author} {\bibinfo {author} {\bibfnamefont {Roman}\ \bibnamefont
  {Bielski}}\ and\ \bibinfo {author} {\bibfnamefont {Michal}\ \bibnamefont
  {Tencer}},\ }\bibfield  {title} {\enquote {\bibinfo {title} {Absolute
  enantioselective separation: Optical activity ex machina},}\ }\href {\doibase
  https://doi.org/10.1002/jssc.200500173} {\bibfield  {journal} {\bibinfo
  {journal} {J. Sep. Sci.}\ }\textbf {\bibinfo {volume} {28}},\ \bibinfo
  {pages} {2325--2332} (\bibinfo {year} {2005})}\BibitemShut {NoStop}%
\bibitem [{\citenamefont {Ye}\ \emph {et~al.}(2019)\citenamefont {Ye},
  \citenamefont {Zhang}, \citenamefont {Chen},\ and\ \citenamefont
  {Li}}]{Intro-select03}%
  \BibitemOpen
  \bibfield  {author} {\bibinfo {author} {\bibfnamefont {Chong}\ \bibnamefont
  {Ye}}, \bibinfo {author} {\bibfnamefont {Quansheng}\ \bibnamefont {Zhang}},
  \bibinfo {author} {\bibfnamefont {Yu-Yuan}\ \bibnamefont {Chen}}, \ and\
  \bibinfo {author} {\bibfnamefont {Yong}\ \bibnamefont {Li}},\ }\bibfield
  {title} {\enquote {\bibinfo {title} {Determination of enantiomeric excess
  with chirality-dependent ac stark effects in cyclic three-level models},}\
  }\href {\doibase 10.1103/PhysRevA.100.033411} {\bibfield  {journal} {\bibinfo
   {journal} {Phys. Rev. A}\ }\textbf {\bibinfo {volume} {100}},\ \bibinfo
  {pages} {033411} (\bibinfo {year} {2019})}\BibitemShut {NoStop}%
\bibitem [{\citenamefont {McKendry}\ \emph {et~al.}(1998)\citenamefont
  {McKendry}, \citenamefont {Theoclitou}, \citenamefont {Rayment},\ and\
  \citenamefont {Abell}}]{Intro-select06}%
  \BibitemOpen
  \bibfield  {author} {\bibinfo {author} {\bibfnamefont {Rachel}\ \bibnamefont
  {McKendry}}, \bibinfo {author} {\bibfnamefont {Maria-Elena}\ \bibnamefont
  {Theoclitou}}, \bibinfo {author} {\bibfnamefont {Trevor}\ \bibnamefont
  {Rayment}}, \ and\ \bibinfo {author} {\bibfnamefont {Chris}\ \bibnamefont
  {Abell}},\ }\bibfield  {title} {\enquote {\bibinfo {title} {Chiral
  discrimination by chemical force microscopy},}\ }\href {\doibase
  10.1038/35339} {\bibfield  {journal} {\bibinfo  {journal} {Nature}\ }\textbf
  {\bibinfo {volume} {391}},\ \bibinfo {pages} {566--568} (\bibinfo {year}
  {1998})}\BibitemShut {NoStop}%
\bibitem [{\citenamefont {Bielski}\ and\ \citenamefont
  {Tencer}(2007)}]{Intro-select07}%
  \BibitemOpen
  \bibfield  {author} {\bibinfo {author} {\bibfnamefont {Roman}\ \bibnamefont
  {Bielski}}\ and\ \bibinfo {author} {\bibfnamefont {Michal}\ \bibnamefont
  {Tencer}},\ }\bibfield  {title} {\enquote {\bibinfo {title} {A possible path
  to the rna world: Enantioselective and diastereoselective purification of
  ribose},}\ }\href {\doibase 10.1007/s11084-006-9022-9} {\bibfield  {journal}
  {\bibinfo  {journal} {Orig Life Evol. Biosph}\ }\textbf {\bibinfo {volume}
  {37}},\ \bibinfo {pages} {167--175} (\bibinfo {year} {2007})}\BibitemShut
  {NoStop}%
\bibitem [{\citenamefont {Kr\'al}\ and\ \citenamefont
  {Shapiro}(2001)}]{Intro-threelevel03}%
  \BibitemOpen
  \bibfield  {author} {\bibinfo {author} {\bibfnamefont {Petr}\ \bibnamefont
  {Kr\'al}}\ and\ \bibinfo {author} {\bibfnamefont {Moshe}\ \bibnamefont
  {Shapiro}},\ }\bibfield  {title} {\enquote {\bibinfo {title} {Cyclic
  population transfer in quantum systems with broken symmetry},}\ }\href
  {\doibase 10.1103/PhysRevLett.87.183002} {\bibfield  {journal} {\bibinfo
  {journal} {Phys. Rev. Lett.}\ }\textbf {\bibinfo {volume} {87}},\ \bibinfo
  {pages} {183002} (\bibinfo {year} {2001})}\BibitemShut {NoStop}%
\bibitem [{\citenamefont {Ye}\ \emph {et~al.}(2018)\citenamefont {Ye},
  \citenamefont {Zhang},\ and\ \citenamefont {Li}}]{ham02}%
  \BibitemOpen
  \bibfield  {author} {\bibinfo {author} {\bibfnamefont {Chong}\ \bibnamefont
  {Ye}}, \bibinfo {author} {\bibfnamefont {Quansheng}\ \bibnamefont {Zhang}}, \
  and\ \bibinfo {author} {\bibfnamefont {Yong}\ \bibnamefont {Li}},\ }\bibfield
   {title} {\enquote {\bibinfo {title} {Real single-loop cyclic three-level
  configuration of chiral molecules},}\ }\href {\doibase
  10.1103/PhysRevA.98.063401} {\bibfield  {journal} {\bibinfo  {journal} {Phys.
  Rev. A}\ }\textbf {\bibinfo {volume} {98}},\ \bibinfo {pages} {063401}
  (\bibinfo {year} {2018})}\BibitemShut {NoStop}%
\bibitem [{\citenamefont {You}\ and\ \citenamefont
  {Nori}(2011)}]{Intro-cycle06}%
  \BibitemOpen
  \bibfield  {author} {\bibinfo {author} {\bibfnamefont {J.~Q.}\ \bibnamefont
  {You}}\ and\ \bibinfo {author} {\bibfnamefont {Franco}\ \bibnamefont
  {Nori}},\ }\bibfield  {title} {\enquote {\bibinfo {title} {Atomic physics and
  quantum optics using superconducting circuits},}\ }\href {\doibase
  10.1038/nature10122} {\bibfield  {journal} {\bibinfo  {journal} {Nature}\
  }\textbf {\bibinfo {volume} {474}},\ \bibinfo {pages} {589--597} (\bibinfo
  {year} {2011})}\BibitemShut {NoStop}%
\bibitem [{\citenamefont {Shapiro}\ and\ \citenamefont
  {Brumer}(1991)}]{Intro-threelevel01}%
  \BibitemOpen
  \bibfield  {author} {\bibinfo {author} {\bibfnamefont {Moshe}\ \bibnamefont
  {Shapiro}}\ and\ \bibinfo {author} {\bibfnamefont {Paul}\ \bibnamefont
  {Brumer}},\ }\bibfield  {title} {\enquote {\bibinfo {title} {Controlled
  photon induced symmetry breaking: Chiral molecular products from achiral
  precursors},}\ }\href {\doibase 10.1063/1.461247} {\bibfield  {journal}
  {\bibinfo  {journal} {J. Chem. Phys.}\ }\textbf {\bibinfo {volume} {95}},\
  \bibinfo {pages} {8658--8661} (\bibinfo {year} {1991})}\BibitemShut {NoStop}%
\bibitem [{\citenamefont {Shapiro}\ \emph {et~al.}(2000)\citenamefont
  {Shapiro}, \citenamefont {Frishman},\ and\ \citenamefont
  {Brumer}}]{Intro-threelevel02}%
  \BibitemOpen
  \bibfield  {author} {\bibinfo {author} {\bibfnamefont {Moshe}\ \bibnamefont
  {Shapiro}}, \bibinfo {author} {\bibfnamefont {Einat}\ \bibnamefont
  {Frishman}}, \ and\ \bibinfo {author} {\bibfnamefont {Paul}\ \bibnamefont
  {Brumer}},\ }\bibfield  {title} {\enquote {\bibinfo {title} {Coherently
  controlled asymmetric synthesis with achiral light},}\ }\href {\doibase
  10.1103/PhysRevLett.84.1669} {\bibfield  {journal} {\bibinfo  {journal}
  {Phys. Rev. Lett.}\ }\textbf {\bibinfo {volume} {84}},\ \bibinfo {pages}
  {1669--1672} (\bibinfo {year} {2000})}\BibitemShut {NoStop}%
\bibitem [{\citenamefont {Li}\ and\ \citenamefont
  {Shapiro}(2010)}]{Intro-threelevel04}%
  \BibitemOpen
  \bibfield  {author} {\bibinfo {author} {\bibfnamefont {Xuan}\ \bibnamefont
  {Li}}\ and\ \bibinfo {author} {\bibfnamefont {Moshe}\ \bibnamefont
  {Shapiro}},\ }\bibfield  {title} {\enquote {\bibinfo {title} {Theory of the
  optical spatial separation of racemic mixtures of chiral molecules},}\ }\href
  {\doibase 10.1063/1.3429884} {\bibfield  {journal} {\bibinfo  {journal} {J.
  Chem. Phys.}\ }\textbf {\bibinfo {volume} {132}},\ \bibinfo {pages} {194315}
  (\bibinfo {year} {2010})}\BibitemShut {NoStop}%
\bibitem [{\citenamefont {Jia}\ and\ \citenamefont
  {Wei}(2011)}]{Intro-threelevel05}%
  \BibitemOpen
  \bibfield  {author} {\bibinfo {author} {\bibfnamefont {W.~Z.}\ \bibnamefont
  {Jia}}\ and\ \bibinfo {author} {\bibfnamefont {L.~F.}\ \bibnamefont {Wei}},\
  }\bibfield  {title} {\enquote {\bibinfo {title} {Probing molecular chirality
  by coherent optical absorption spectra},}\ }\href {\doibase
  10.1103/PhysRevA.84.053849} {\bibfield  {journal} {\bibinfo  {journal} {Phys.
  Rev. A}\ }\textbf {\bibinfo {volume} {84}},\ \bibinfo {pages} {053849}
  (\bibinfo {year} {2011})}\BibitemShut {NoStop}%
\bibitem [{\citenamefont {Patterson}\ \emph {et~al.}(2013)\citenamefont
  {Patterson}, \citenamefont {Schnell},\ and\ \citenamefont
  {Doyle}}]{Intro-threelevel06}%
  \BibitemOpen
  \bibfield  {author} {\bibinfo {author} {\bibfnamefont {David}\ \bibnamefont
  {Patterson}}, \bibinfo {author} {\bibfnamefont {Melanie}\ \bibnamefont
  {Schnell}}, \ and\ \bibinfo {author} {\bibfnamefont {John~M.}\ \bibnamefont
  {Doyle}},\ }\bibfield  {title} {\enquote {\bibinfo {title}
  {Enantiomer-specific detection of chiral molecules via microwave
  spectroscopy},}\ }\href {\doibase 10.1038/nature12150} {\bibfield  {journal}
  {\bibinfo  {journal} {Nature}\ }\textbf {\bibinfo {volume} {497}},\ \bibinfo
  {pages} {475--477} (\bibinfo {year} {2013})}\BibitemShut {NoStop}%
\bibitem [{\citenamefont {Patterson}\ and\ \citenamefont
  {Doyle}(2013)}]{Intro-threelevel07}%
  \BibitemOpen
  \bibfield  {author} {\bibinfo {author} {\bibfnamefont {David}\ \bibnamefont
  {Patterson}}\ and\ \bibinfo {author} {\bibfnamefont {John~M.}\ \bibnamefont
  {Doyle}},\ }\bibfield  {title} {\enquote {\bibinfo {title} {Sensitive chiral
  analysis via microwave three-wave mixing},}\ }\href {\doibase
  10.1103/PhysRevLett.111.023008} {\bibfield  {journal} {\bibinfo  {journal}
  {Phys. Rev. Lett.}\ }\textbf {\bibinfo {volume} {111}},\ \bibinfo {pages}
  {023008} (\bibinfo {year} {2013})}\BibitemShut {NoStop}%
\bibitem [{\citenamefont {Orlando}\ \emph {et~al.}(1999)\citenamefont
  {Orlando}, \citenamefont {Mooij}, \citenamefont {Tian}, \citenamefont
  {van~der Wal}, \citenamefont {Levitov}, \citenamefont {Lloyd},\ and\
  \citenamefont {Mazo}}]{Intro-cycle01}%
  \BibitemOpen
  \bibfield  {author} {\bibinfo {author} {\bibfnamefont {T.~P.}\ \bibnamefont
  {Orlando}}, \bibinfo {author} {\bibfnamefont {J.~E.}\ \bibnamefont {Mooij}},
  \bibinfo {author} {\bibfnamefont {Lin}\ \bibnamefont {Tian}}, \bibinfo
  {author} {\bibfnamefont {Caspar~H.}\ \bibnamefont {van~der Wal}}, \bibinfo
  {author} {\bibfnamefont {L.~S.}\ \bibnamefont {Levitov}}, \bibinfo {author}
  {\bibfnamefont {Seth}\ \bibnamefont {Lloyd}}, \ and\ \bibinfo {author}
  {\bibfnamefont {J.~J.}\ \bibnamefont {Mazo}},\ }\bibfield  {title} {\enquote
  {\bibinfo {title} {Superconducting persistent-current qubit},}\ }\href
  {\doibase 10.1103/PhysRevB.60.15398} {\bibfield  {journal} {\bibinfo
  {journal} {Phys. Rev. B}\ }\textbf {\bibinfo {volume} {60}},\ \bibinfo
  {pages} {15398--15413} (\bibinfo {year} {1999})}\BibitemShut {NoStop}%
\bibitem [{\citenamefont {H\"onigl-Decrinis}\ \emph {et~al.}(2018)\citenamefont
  {H\"onigl-Decrinis}, \citenamefont {Antonov}, \citenamefont {Shaikhaidarov},
  \citenamefont {Antonov}, \citenamefont {Dmitriev},\ and\ \citenamefont
  {Astafiev}}]{Intro-cycle02}%
  \BibitemOpen
  \bibfield  {author} {\bibinfo {author} {\bibfnamefont {T.}~\bibnamefont
  {H\"onigl-Decrinis}}, \bibinfo {author} {\bibfnamefont {I.~V.}\ \bibnamefont
  {Antonov}}, \bibinfo {author} {\bibfnamefont {R.}~\bibnamefont
  {Shaikhaidarov}}, \bibinfo {author} {\bibfnamefont {V.~N.}\ \bibnamefont
  {Antonov}}, \bibinfo {author} {\bibfnamefont {A.~Yu.}\ \bibnamefont
  {Dmitriev}}, \ and\ \bibinfo {author} {\bibfnamefont {O.~V.}\ \bibnamefont
  {Astafiev}},\ }\bibfield  {title} {\enquote {\bibinfo {title} {Mixing of
  coherent waves in a single three-level artificial atom},}\ }\href {\doibase
  10.1103/PhysRevA.98.041801} {\bibfield  {journal} {\bibinfo  {journal} {Phys.
  Rev. A}\ }\textbf {\bibinfo {volume} {98}},\ \bibinfo {pages} {041801}
  (\bibinfo {year} {2018})}\BibitemShut {NoStop}%
\bibitem [{\citenamefont {Guo}\ \emph {et~al.}(2022)\citenamefont {Guo},
  \citenamefont {Gong}, \citenamefont {Ma},\ and\ \citenamefont
  {Shu}}]{Intro-cycle03}%
  \BibitemOpen
  \bibfield  {author} {\bibinfo {author} {\bibfnamefont {Yu}~\bibnamefont
  {Guo}}, \bibinfo {author} {\bibfnamefont {Xun}\ \bibnamefont {Gong}},
  \bibinfo {author} {\bibfnamefont {Songshan}\ \bibnamefont {Ma}}, \ and\
  \bibinfo {author} {\bibfnamefont {Chuan-Cun}\ \bibnamefont {Shu}},\
  }\bibfield  {title} {\enquote {\bibinfo {title} {Cyclic
  three-level-pulse-area theorem for enantioselective state transfer of chiral
  molecules},}\ }\href {\doibase 10.1103/PhysRevA.105.013102} {\bibfield
  {journal} {\bibinfo  {journal} {Phys. Rev. A}\ }\textbf {\bibinfo {volume}
  {105}},\ \bibinfo {pages} {013102} (\bibinfo {year} {2022})}\BibitemShut
  {NoStop}%
\bibitem [{\citenamefont {Leibscher}\ \emph {et~al.}(2019)\citenamefont
  {Leibscher}, \citenamefont {Giesen},\ and\ \citenamefont {Koch}}]{chiral001}%
  \BibitemOpen
  \bibfield  {author} {\bibinfo {author} {\bibfnamefont {Monika}\ \bibnamefont
  {Leibscher}}, \bibinfo {author} {\bibfnamefont {Thomas~F.}\ \bibnamefont
  {Giesen}}, \ and\ \bibinfo {author} {\bibfnamefont {Christiane~P.}\
  \bibnamefont {Koch}},\ }\bibfield  {title} {\enquote {\bibinfo {title}
  {Principles of enantio-selective excitation in three-wave mixing spectroscopy
  of chiral molecules},}\ }\href {\doibase 10.1063/1.5097406} {\bibfield
  {journal} {\bibinfo  {journal} {J. Chem. Phys.}\ }\textbf {\bibinfo {volume}
  {151}},\ \bibinfo {pages} {014302} (\bibinfo {year} {2019})}\BibitemShut
  {NoStop}%
\bibitem [{\citenamefont {Quan}\ \emph {et~al.}(2006)\citenamefont {Quan},
  \citenamefont {Wang}, \citenamefont {Liu}, \citenamefont {Sun},\ and\
  \citenamefont {Nori}}]{QHE001}%
  \BibitemOpen
  \bibfield  {author} {\bibinfo {author} {\bibfnamefont {H.~T.}\ \bibnamefont
  {Quan}}, \bibinfo {author} {\bibfnamefont {Y.~D.}\ \bibnamefont {Wang}},
  \bibinfo {author} {\bibfnamefont {Yu-xi}\ \bibnamefont {Liu}}, \bibinfo
  {author} {\bibfnamefont {C.~P.}\ \bibnamefont {Sun}}, \ and\ \bibinfo
  {author} {\bibfnamefont {Franco}\ \bibnamefont {Nori}},\ }\bibfield  {title}
  {\enquote {\bibinfo {title} {Maxwell's demon assisted thermodynamic cycle in
  superconducting quantum circuits},}\ }\href {\doibase
  10.1103/PhysRevLett.97.180402} {\bibfield  {journal} {\bibinfo  {journal}
  {Phys. Rev. Lett.}\ }\textbf {\bibinfo {volume} {97}},\ \bibinfo {pages}
  {180402} (\bibinfo {year} {2006})}\BibitemShut {NoStop}%
\bibitem [{\citenamefont {Quan}(2009)}]{QHE002}%
  \BibitemOpen
  \bibfield  {author} {\bibinfo {author} {\bibfnamefont {H.~T.}\ \bibnamefont
  {Quan}},\ }\bibfield  {title} {\enquote {\bibinfo {title} {Quantum
  thermodynamic cycles and quantum heat engines. ii.}}\ }\href {\doibase
  10.1103/PhysRevE.79.041129} {\bibfield  {journal} {\bibinfo  {journal} {Phys.
  Rev. E}\ }\textbf {\bibinfo {volume} {79}},\ \bibinfo {pages} {041129}
  (\bibinfo {year} {2009})}\BibitemShut {NoStop}%
\bibitem [{\citenamefont {Hildred}\ \emph {et~al.}(1983)\citenamefont
  {Hildred}, \citenamefont {Hassan}, \citenamefont {Puri},\ and\ \citenamefont
  {Bullough}}]{thermalization001}%
  \BibitemOpen
  \bibfield  {author} {\bibinfo {author} {\bibfnamefont {G~P}\ \bibnamefont
  {Hildred}}, \bibinfo {author} {\bibfnamefont {S~S}\ \bibnamefont {Hassan}},
  \bibinfo {author} {\bibfnamefont {R~R}\ \bibnamefont {Puri}}, \ and\ \bibinfo
  {author} {\bibfnamefont {R~K}\ \bibnamefont {Bullough}},\ }\bibfield  {title}
  {\enquote {\bibinfo {title} {Thermal reservoir effects in resonance
  fluorescence},}\ }\href {\doibase 10.1088/0022-3700/16/10/008} {\bibfield
  {journal} {\bibinfo  {journal} {J. Phys. B: Atom. Mol. Phys.}\ }\textbf
  {\bibinfo {volume} {16}},\ \bibinfo {pages} {1703} (\bibinfo {year}
  {1983})}\BibitemShut {NoStop}%
\bibitem [{\citenamefont {Breuer}\ and\ \citenamefont
  {Petruccione}(2007)}]{QopenSystemBook}%
  \BibitemOpen
  \bibfield  {author} {\bibinfo {author} {\bibfnamefont {Heinz-Peter}\
  \bibnamefont {Breuer}}\ and\ \bibinfo {author} {\bibfnamefont {Francesco}\
  \bibnamefont {Petruccione}},\ }\href {\doibase
  10.1093/acprof:oso/9780199213900.001.0001} {\emph {\bibinfo {title} {{The
  Theory of Open Quantum Systems}}}}\ (\bibinfo  {publisher} {Oxford University
  Press},\ \bibinfo {year} {2007})\BibitemShut {NoStop}%
\bibitem [{\citenamefont {Scully}\ and\ \citenamefont
  {Zubairy}(1997)}]{QOscully}%
  \BibitemOpen
  \bibfield  {author} {\bibinfo {author} {\bibfnamefont {Marlan~O.}\
  \bibnamefont {Scully}}\ and\ \bibinfo {author} {\bibfnamefont {M.~Suhail}\
  \bibnamefont {Zubairy}},\ }\href {\doibase 10.1017/CBO9780511813993} {\emph
  {\bibinfo {title} {Quantum Optics}}}\ (\bibinfo  {publisher} {Cambridge
  University Press},\ \bibinfo {year} {1997})\BibitemShut {NoStop}%
\bibitem [{\citenamefont {Fink}\ \emph {et~al.}(2010)\citenamefont {Fink},
  \citenamefont {Steffen}, \citenamefont {Studer}, \citenamefont {Bishop},
  \citenamefont {Baur}, \citenamefont {Bianchetti}, \citenamefont {Bozyigit},
  \citenamefont {Lang}, \citenamefont {Filipp}, \citenamefont {Leek},\ and\
  \citenamefont {Wallraff}}]{localMasterEq}%
  \BibitemOpen
  \bibfield  {author} {\bibinfo {author} {\bibfnamefont {J.~M.}\ \bibnamefont
  {Fink}}, \bibinfo {author} {\bibfnamefont {L.}~\bibnamefont {Steffen}},
  \bibinfo {author} {\bibfnamefont {P.}~\bibnamefont {Studer}}, \bibinfo
  {author} {\bibfnamefont {Lev~S.}\ \bibnamefont {Bishop}}, \bibinfo {author}
  {\bibfnamefont {M.}~\bibnamefont {Baur}}, \bibinfo {author} {\bibfnamefont
  {R.}~\bibnamefont {Bianchetti}}, \bibinfo {author} {\bibfnamefont
  {D.}~\bibnamefont {Bozyigit}}, \bibinfo {author} {\bibfnamefont
  {C.}~\bibnamefont {Lang}}, \bibinfo {author} {\bibfnamefont {S.}~\bibnamefont
  {Filipp}}, \bibinfo {author} {\bibfnamefont {P.~J.}\ \bibnamefont {Leek}}, \
  and\ \bibinfo {author} {\bibfnamefont {A.}~\bibnamefont {Wallraff}},\
  }\bibfield  {title} {\enquote {\bibinfo {title} {Quantum-to-classical
  transition in cavity quantum electrodynamics},}\ }\href {\doibase
  10.1103/PhysRevLett.105.163601} {\bibfield  {journal} {\bibinfo  {journal}
  {Phys. Rev. Lett.}\ }\textbf {\bibinfo {volume} {105}},\ \bibinfo {pages}
  {163601} (\bibinfo {year} {2010})}\BibitemShut {NoStop}%
\bibitem [{\citenamefont {Johansson}\ \emph {et~al.}(2013)\citenamefont
  {Johansson}, \citenamefont {Nation},\ and\ \citenamefont {Nori}}]{Qutip}%
  \BibitemOpen
  \bibfield  {author} {\bibinfo {author} {\bibfnamefont {J.R.}\ \bibnamefont
  {Johansson}}, \bibinfo {author} {\bibfnamefont {P.D.}\ \bibnamefont
  {Nation}}, \ and\ \bibinfo {author} {\bibfnamefont {Franco}\ \bibnamefont
  {Nori}},\ }\bibfield  {title} {\enquote {\bibinfo {title} {Qutip 2: A python
  framework for the dynamics of open quantum systems},}\ }\href {\doibase
  https://doi.org/10.1016/j.cpc.2012.11.019} {\bibfield  {journal} {\bibinfo
  {journal} {Comput. Phys. Commun.}\ }\textbf {\bibinfo {volume} {184}},\
  \bibinfo {pages} {1234--1240} (\bibinfo {year} {2013})}\BibitemShut {NoStop}%
\bibitem [{\citenamefont {Leggio}\ and\ \citenamefont
  {Antezza}(2016)}]{adiabatic01}%
  \BibitemOpen
  \bibfield  {author} {\bibinfo {author} {\bibfnamefont {Bruno}\ \bibnamefont
  {Leggio}}\ and\ \bibinfo {author} {\bibfnamefont {Mauro}\ \bibnamefont
  {Antezza}},\ }\bibfield  {title} {\enquote {\bibinfo {title} {Otto engine
  beyond its standard quantum limit},}\ }\href {\doibase
  10.1103/PhysRevE.93.022122} {\bibfield  {journal} {\bibinfo  {journal} {Phys.
  Rev. E}\ }\textbf {\bibinfo {volume} {93}},\ \bibinfo {pages} {022122}
  (\bibinfo {year} {2016})}\BibitemShut {NoStop}%
\bibitem [{\citenamefont {Solfanelli}\ \emph {et~al.}(2020)\citenamefont
  {Solfanelli}, \citenamefont {Falsetti},\ and\ \citenamefont
  {Campisi}}]{thermalMachines001}%
  \BibitemOpen
  \bibfield  {author} {\bibinfo {author} {\bibfnamefont {Andrea}\ \bibnamefont
  {Solfanelli}}, \bibinfo {author} {\bibfnamefont {Marco}\ \bibnamefont
  {Falsetti}}, \ and\ \bibinfo {author} {\bibfnamefont {Michele}\ \bibnamefont
  {Campisi}},\ }\bibfield  {title} {\enquote {\bibinfo {title} {Nonadiabatic
  single-qubit quantum otto engine},}\ }\href {\doibase
  10.1103/PhysRevB.101.054513} {\bibfield  {journal} {\bibinfo  {journal}
  {Phys. Rev. B}\ }\textbf {\bibinfo {volume} {101}},\ \bibinfo {pages}
  {054513} (\bibinfo {year} {2020})}\BibitemShut {NoStop}%
\bibitem [{\citenamefont {Quack}(1986)}]{QUACK1986147}%
  \BibitemOpen
  \bibfield  {author} {\bibinfo {author} {\bibfnamefont {Martin}\ \bibnamefont
  {Quack}},\ }\bibfield  {title} {\enquote {\bibinfo {title} {On the
  measurement of the parity violating energy difference between enantiomers},}\
  }\href {\doibase https://doi.org/10.1016/0009-2614(86)80098-7} {\bibfield
  {journal} {\bibinfo  {journal} {Chem. Phys. Lett.}\ }\textbf {\bibinfo
  {volume} {132}},\ \bibinfo {pages} {147--153} (\bibinfo {year}
  {1986})}\BibitemShut {NoStop}%
\bibitem [{\citenamefont {Fábri}\ \emph {et~al.}(2015)\citenamefont {Fábri},
  \citenamefont {Horný},\ and\ \citenamefont {Quack}}]{cphc.201500801}%
  \BibitemOpen
  \bibfield  {author} {\bibinfo {author} {\bibfnamefont {Csaba}\ \bibnamefont
  {Fábri}}, \bibinfo {author} {\bibfnamefont {Ľuboš}\ \bibnamefont
  {Horný}}, \ and\ \bibinfo {author} {\bibfnamefont {Martin}\ \bibnamefont
  {Quack}},\ }\bibfield  {title} {\enquote {\bibinfo {title} {Tunneling and
  parity violation in trisulfane (hsssh): An almost ideal molecule for
  detecting parity violation in chiral molecules},}\ }\href {\doibase
  https://doi.org/10.1002/cphc.201500801} {\bibfield  {journal} {\bibinfo
  {journal} {ChemPhysChem}\ }\textbf {\bibinfo {volume} {16}},\ \bibinfo
  {pages} {3584--3589} (\bibinfo {year} {2015})}\BibitemShut {NoStop}%
\end{thebibliography}%
\end{document}